\documentclass[12pt,draftcls,peerreview,onecolumn]{IEEEtran}

\usepackage{theorem}
\usepackage{times}
\usepackage[dvips]{graphicx,color}
\usepackage{exscale}
\usepackage{epic,eepic}
\usepackage{amsmath}
\usepackage{amssymb}
\usepackage{cite, verbatim}
\usepackage{psfrag}

\newtheorem{definition}{Definition}
\newtheorem{theorem}{Theorem}

\newcommand{\qed}{\nobreak \ifvmode \relax \else
      \ifdim\lastskip<1.5em \hskip-\lastskip
      \hskip1.5em plus0em minus0.5em \fi \nobreak
      \vrule height0.75em width0.5em depth0.25em\fi}

\begin{document}

\title{Source-Channel Coding and Separation for Generalized Communication Systems}
\author{Yifan~Liang,~\IEEEmembership{Student~Member,~IEEE},
        Andrea~Goldsmith,~\IEEEmembership{Fellow,~IEEE,}
        and~Michelle~Effros,~\IEEEmembership{Senior~Member,~IEEE}
\thanks{
This work was supported by the DARPA ITMANET program under grant
number 1105741-1-TFIND. The material in this paper was presented in
part at the IEEE Information Theory Workshop, Lake Tahoe,
California, September 2007.}
\thanks{Y. Liang and A. Goldsmith are with the Department of~Electrical Engineering, Stanford
University, Stanford CA 94305 (email: yfl@wsl.stanford.edu,
andrea@wsl.stanford.edu). }
\thanks{M. Effros is with the Department
of Electrical Engineering, California Institute of Technology,
Pasadena, CA 91125 (email: effros@caltech.edu).}
}

\maketitle

\newcommand{\beq}{\begin{equation}}
\newcommand{\eeq}{\end{equation}}
\newcommand{\ubar}{\underline}
\newcommand{\typ}{{A_\epsilon^{(n)}}}
\newcommand{\Pens}{P_e^{(n,s)}}
\newcommand{\bs}[1]{\boldsymbol{#1}}
\newcommand{\Fig}[2]
{\begin{figure}[htbp]
\begin{center}
\includegraphics[width=3in]{EPS/#1.eps}
\caption{#2} \label{fig:#1}
\end{center}
\end{figure}}
\newcommand{\refE}[1] {Eqn (\ref{eqn:#1})}
\newcommand{\refS}[1] {Sec. \ref{sec:#1}}
\newcommand{\refF}[1] {Fig. \ref{fig:#1}}
\newcommand{\bsi}[1] {b_s^{(#1)}}
\newcommand{\bpi}[1] {b_p^{(#1)}}
\newcommand{\BC} {\textrm{BC}}

\begin{abstract}
We consider transmission of stationary and ergodic sources over
non-ergodic composite channels with channel state information at the
receiver (CSIR). Previously we introduced alternate capacity
definitions to Shannon capacity, including the capacity versus
outage and the expected capacity. These generalized definitions
relax the constraint of Shannon capacity that all transmitted
information must be decoded at the receiver. In this work alternate
end-to-end distortion metrics such as the distortion versus outage
and the expected distortion are introduced to relax the constraint
that a single distortion level has to be maintained for all channel
states. For transmission of stationary and ergodic sources over
stationary and ergodic channels, the classical Shannon separation
theorem enables separate design of source and channel codes and
guarantees optimal performance. For generalized communication
systems, we show that different end-to-end distortion metrics lead
to different conclusions about separation optimality even for the
same source and channel models.

Separation does not imply isolation - the source and channel still
need to communicate with each other through some interfaces. For
Shannon separation schemes, the interface is a single-number
comparison between the source coding rate and the channel capacity.
Here we include a broader class of transmission schemes as
separation schemes by relaxing the constraint of a single-number
interface. We show that one such generalized scheme guarantees the
separation optimality under the distortion versus outage metric.
Under the expected distortion metric, separation schemes are no
longer optimal.  We expect a performance enhancement when the source
and channel coders exchange more information through more
sophisticated interfaces, and illustrate the tradeoff between
interface complexity and end-to-end performance through the example
of transmitting a binary symmetric source over a composite binary
symmetric channel.
\end{abstract}


\IEEEpeerreviewmaketitle

\section{Introduction}
The time-varying nature of the underlying channel is one of the most
significant design challenges in wireless communication systems. In
particular, real-time media traffic typically has a stringent delay
constraint, so the exploitation of long blocklength frames is
infeasible and the entire frame may fall into deep fading channel
states. Furthermore, the receiver may have limited resources to feed
the estimated channel state information back to the transmitter,
which precludes adaptive transmission and forces the transmitter to
use a stationary coding strategy. The above described situation is
modeled as a slowly fading channel with receiver side information
only, which is an example of a
%
non-ergodic {\em composite channel}. A composite channel is a
collection of component channels $\{W_S:S\in{\cal S}\}$
parameterized by $S$, where the random variable $S$ is chosen
according to some distribution $p(S)$ at the beginning of
transmission and then held fixed. We assume the channel realization
is revealed to the receiver but not the transmitter. This class of
channel is also referred to as the {\em mixed channel}
\cite{HanBook} or the {\em averaged channel} \cite{Ahlswede68} in
literature.

The Shannon capacity of a composite channel is given by the
Verd\'u-Han generalized capacity formula \cite{VerduH:94}
\[
C = \sup_{\bs{X}} \underline{\bs{I}} (\bs{X}; \bs{Y}),
\]
where $\underline{\bs{I}} (\bs{X}; \bs{Y})$ is the liminf in
probability of the normalized information density. This formula
highlights the pessimistic nature of the Shannon capacity
definition, which is dominated by the performance of the ``worst''
channel, no matter how small its probability. To provide more
flexibility in capacity definitions for composite channels, in
\cite{Effros98, LiangJournal} we relax the constraint that all
transmitted information has to be correctly decoded and derive
alternate definitions including the {\em capacity versus outage} and
the {\em expected capacity}. The capacity versus outage approach
allows certain data loss in some channel states in exchange for
higher rates in other states. It was previously examined in
\cite{OzarowS:94} for single-antenna cellular systems, and later
became a common criterion for multiple-antenna wireless fading
channels \cite{FoschiniG:97, MIMOJSAC, Zheng02}. See
\cite[Ch.~4]{Goldsmith} and references therein for more details. The
expected capacity approach also requires the transmitter to use a
single encoder but allows the receiver to choose from a collection
of decoders based on channel states. It was derived for a Gaussian
slow-fading channel in~\cite{Shamai03}, and for a composite binary
symmetric channel (BSC) in \cite{Liang072}.

Channel capacity theorems deal with data transmission in a
communication system. When extending the system to include the
source of the data, we also need to consider the data compression
problem which deals with source representation and reconstruction.
For the overall system, the end-to-end distortion is a well-accepted
performance metric. When both the source and channel are stationary
and ergodic, codes are usually designed to achieve the same
end-to-end distortion level for any source sequence and channel
realization. Nevertheless, practical systems do not always impose
this constraint. If the channel model is generalized to such
scenarios as the composite channel above, it is natural to relax the
constraint that a single distortion level has to be maintained for
all channel states. In parallel with the development of alternative
capacity definitions, we introduce generalized end-to-end distortion
metrics including the {\em distortion versus outage} and the {\em
expected distortion}. The distortion versus outage is characterized
by a pair $(q, D_q)$, where the distortion level $D_q$ is guaranteed
in receiver-recognized non-outage states of probability no less than
$(1-q)$. This definition requires CSIR based on which the outage can
be declared. The expected distortion is defined as
$\mathbb{E}_SD_S$, i.e. the achievable distortion $D_S$ in channel
state $S$ averaged over
the underlying distribution $p(S)$.
%
%
These alternative distortion metrics are also considered in prior
works. In \cite{Honig04} the average distortion $q\sigma^2 + (1-q)
D_q$, obtained by averaging over outage and non-outage states, was
adopted as a fidelity criterion to analyze a two-hop fading channel.
Here $\sigma^2$ is the variance of the source symbols. The expected
distortion was analyzed for the MIMO block fading channel in the
high SNR regime \cite{Gunduz} and in the finite SNR regime
\cite{Ng07a, Ng07b}. Various coding schemes for expected distortion
were also studied in a slightly different but closely related
broadcast scenario \cite{Shamai98, Reznic06, Mittal02}.

Data compression (source coding) and data transmission (channel
coding) are two fundamental topics in Shannon theory. For
transmission of a discrete memoryless source (DMS) over a discrete
memoryless channel (DMC), the renowned source-channel separation
theorem \cite[Theorem 2.4]{CsiszarK:81} asserts that a target
distortion level $D$ is achievable if and only if the channel
capacity $C$ exceeds the source rate distortion function $R(D)$, and
a two-stage separate source-channel code suffices to meet the
requirement\footnote{The separation theorem for lossless
transmission \cite{Shannon48} can be regarded as a special case of
zero distortion.}. This theorem enables separate designs of source
and channel codes with guaranteed optimal performance. It also
extends to stationary and ergodic source and channel models
\cite{Dobrushin63} \cite{Hu64}.
%
Separate source-channel coding schemes provide flexibility through
modularized design. 
From the source's point of view, the source can be transmitted over
any channel with capacity greater than $R(D)$ and be recovered at
the receiver subject to a certain fidelity criterion (the distortion
$D$). The source is indifferent to the statistics of each individual
channel and consequently focuses on source code design independent
of channel statistics.

Despite their flexibility and optimality for certain systems,
separation schemes also have their disadvantages. First of all, the
source encoder needs to observe a long-blocklength source sequence
in order to determine the output, which causes infinite delay.
Second, separation schemes may increase complexity in encoders and
decoders because the two processes of source and channel coding are
acting in opposition to some extent.
Source coding is essentially a data compression process, which
aims at removing redundancy from source sequences to achieve the
most concise representation. 
On the other hand, channel coding deals with data transmission,
which tries to add some redundancy to the transmitted sequence for
robustness against the channel noise. 
If the source redundancy can be exploited by the channel code, then
a joint source-channel coding scheme may avoid this overhead. In
particular, transmission of a Gaussian source over a Gaussian
channel, and a binary symmetric source over a BSC, are both examples
where optimal performance can be achieved without any coding
\cite{Gastpar03}. This is because the source and channel are
``matched" to each other in the sense that the transition
probabilities of the channel solve the variational problem defining
the source rate-distortion function $R(D)$ and the letter
probabilities of the source drive the channel at capacity
\cite[p.74]{Berger}.

A careful inspection of the Shannon separation theorem reveals some
important underlying assumptions: a single-user channel, a
stationary and ergodic source and channel, and a single distortion
level maintained for all transmissions.
Violation of any of these assumptions will likely prompt
reexamination of the separation theorem. For example, Cover et. al.
showed that for a multiple access channel with correlated sources,
the separation theorem fails \cite{ElGamal80}. In \cite{Vembu95}
Vembu et al. gave an example of a non-stationary system where the
source is transmissible through the channel with zero error, yet its
minimum achievable source coding rate is twice the channel capacity.
In this work, we illustrate that different end-to-end distortion
metrics lead to different conclusions about separability even for
the same source and channel model. In fact, source-channel
separation holds under the distortion versus outage metric but fails
under the expected distortion metric. In \cite{Liang07ITW} we proved
the direct part of source-channel separation under the distortion
versus outage metric and established the converse for a system of
Gaussian source and slow-fading Gaussian channels. Here we extend
the converse to more general systems of stationary sources and
composite channels.

Source-channel separation implies that the operation of source and
channel coding does not depend on the statistics of the counterpart.
However, the source and channel do need to communicate with each
other through a {\em negotiation interface} even before the actual
transmission starts. In the classical view of Shannon separation for
stationary ergodic sources and channels, the source requires a rate
$R(D)$ based on the target distortion $D$ and the channel decides if
it can support the rate based on its capacity $C$.
%
For generalized source/channel models and distortion metrics, the
interface is not necessarily a single rate and may allow multiple
parameters to be agreed upon between the source and channel. After
communication through the appropriate negotiation interface, the
source and channel codes may be designed separately and still
achieve the optimal performance.
Vembu et al. studied the transmission of non-stationary sources over
non-stationary channels and
observed that the notion of (strict) domination \cite[Theorem
7]{Vembu95} dictates whether a source is transmissible over a
channel, instead of the simple comparison between the minimum source
coding rate and the channel capacity. The notion of (strict)
domination requires the source to provide the distribution of the
{\em entropy density} and the channel to provide the distribution of
the {\em information density} as the appropriate interface.

The source-channel interface concept also applies after the actual
transmission starts. At the transmitter end, we see examples where
the source sequence is directly supplied to the channel, such as the
uncoded transmission of a Gaussian source over a Gaussian channel.
But more generally there is certain processing on the source side,
and the processed output, instead of the original source sequence,
is supplied to the channel. The {\em transmitter interface} contains
what the source actually delivers to the channel. For example, in
separation schemes the interface is the source encoder output; in
hybrid digital-analog schemes \cite{Mittal02} the interface is a
combination of vector quantizer output and quantization residue.
Similarly we can introduce the concept of a {\em receiver
interface}. Instead of directly delivering the channel output
sequence to the destination, the receiver may implement certain
decoding and choose the channel decoder output as the interface. The
interfaces at the transmitter and the receiver are the same in
classical Shannon separation schemes, since the channel code
requires all transmitted information to be correctly decoded with
vanishing error, but in general the two interfaces can be different.
For example, the receiver interface may include an outage indicator
or partial decoding when considering generalized capacity
definitions.

Different transmission schemes can be compared by their end-to-end
performance. Nevertheless, the concept of source-channel interface
opens a new dimension for comparison. Ideally the interface
complexity should be measured by some quantified metrics.
Transmission schemes with low interface complexity are also
appealing in view of simplified system design. We expect a
performance enhancement when the source and channel exchange more
information through a more sophisticated interface, and illustrate
the tradeoff between interface complexity and end-to-end performance
through some examples in this work.


The rest of the paper is organized as follows. We review alternative
channel capacity definitions and define corresponding end-to-end
distortion metrics in Section \ref{sec:performance}.
In Section \ref{sec:source-channel} we provide a new perspective of
source-channel separation generalized from Shannon's classical view
and also introduce the concept of source-channel interface. In
Section \ref{sec:outageDistortion} we establish the separation
optimality for transmission of stationary ergodic sources over
composite channels under the distortion versus outage metric. In
Section \ref{sec:InterfaceBSC} we consider various schemes to
transmit a binary symmetric source (BSS) over a composite BSC and
show the tradeoff between achievable expected distortion and
interface complexity. Conclusions are given in Section
\ref{sec:con}.

\section{Generalized Performance Metrics}
\label{sec:performance} We first review alternate channel capacity
definitions derived in \cite{Effros98, Liang072} to provide some
background information. We then define alternate end-to-end
performance metrics for the entire communication system, including
the source and the destination.
\subsection{Background: Channel Capacity
Metrics} \label{sec:channelMetrics}  The channel $\bs{W}$ is
statistically modeled as a sequence of $n$-dimensional conditional
distributions $\bs{W}=\{W^n=P_{Z^n|X^n}\}_{n=1}^\infty$. For any
integer $n$, $W^n$ is the conditional distribution from the input
space ${\cal X}^n$ to the output space ${\cal Z}^n$. Let $\bs{X}$
and $\bs{Z}$ denote the input and output processes, respectively.
Each process is specified by a sequence of finite-dimensional
distributions, e.g.
$\bs{X}=\{X^n=(X_1^{(n)},\cdots,X_n^{(n)})\}_{n=1}^\infty$.

In a composite channel, when the channel side information is
available at the receiver, we represent it as an additional channel
output. Specifically, we let $Z^n=(S,Y^n)$, where $S$ is the channel
side information and $Y^n$ is the output of the channel described by
parameter $S$. Throughout, we assume the random variable $S$ is
independent of $\bs{X}$ and unknown to the encoder. Thus for each
$n$
\begin{eqnarray}
P_{W^n}(z^n|x^n) &=& P_{Z^n|X^n}(s, y^n|x^n) \nonumber \\
&=& P_{S}(s) P_{Y^n|X^n,S}(y^n|x^n,s). \label{eqn:CompositeChan}
\end{eqnarray}
The information density is defined similarly as in \cite{VerduH:94}
\begin{eqnarray}
i_{X^nW^n}(x^n;z^n)
&=& \log\frac{P_{W^n}(z^n|x^n)}{P_{Z^n}(z^n)} \nonumber \\
&=& \log\frac{P_{Y^n|X^n,S}(y^n|x^n,s)}{P_{Y^n|S}(y^n|s)} \nonumber
\\
&=& i_{X^nW^n}(x^n;y^n|s). \label{eqn:infoden}
\end{eqnarray}

\subsubsection{Capacity versus Outage} Consider a sequence of $(n,2^{nR})$
codes. Let $P_o^{(n)}$ be the probability that the receiver declares
an outage, and $P_e^{(n)}$ be the decoding error probability
given that no outage is declared. We
say that a rate $R$ is outage-$q$ achievable if there exists a
sequence of $(n,2^{nR})$ channel codes such that ${\displaystyle
\lim_{n\rightarrow\infty}P_o^{(n)}\leq q }$ and ${\displaystyle
\lim_{n\rightarrow\infty}P_e^{(n)} = 0}$. The {\em capacity versus
outage} $C_q$ 
is defined to be the supremum
over all outage-$q$ achievable rates, and is shown to be
\cite{VerduH:94, Effros98}
\begin{equation}
C_q  = \sup_{\bs{X}} \sup \left\{ \alpha:
        \lim_{n\rightarrow\infty} \Pr
        \left[ \frac{1}{n}
        i(X^n;Y^n|S) \leq \alpha \right] \leq q \right\}.
\label{eqn:outage}
\end{equation}
%

The operational implication of this definition is that the encoder
uses a single codebook and sends information at a fixed rate $C_q$.
Assuming repeated channel use and independent channel state at each
use, the receiver can correctly decode the information a proportion
$(1-q)$ of the time and turn itself off a proportion $q$ of the
time. We further define the {\em outage capacity} $C^o_q = (1-q)C_q$
as the long-term average rate, which is a meaningful metric if we
are only interested in the fraction of correctly received packets
and approximate the unreliable packets by surrounding samples, or if
there is some repetition mechanism where the receiver requests
retransmission of lost information from the sender. The value $q$
can be chosen to maximize the long-term average throughput $C^o_q$.


\subsubsection{Expected Capacity}
This notion provides another strategy for increasing
reliably-received rate. Although the transmitter is forced to use a
single encoder at a rate $R_t$ without channel state information,
the receiver can choose from a collection of decoders, each
parameterized by $s$ and decoding at a rate $R_s \le R_t$, based on
CSIR. Denote by $P_e^{(n,s)}$ the error probability associated with
channel state $s$. The expected capacity $C^e$ is the supremum of
all achievable rates $\mathbb{E}_SR_S$ of any code sequence that has
$\mathbb{E}_S P_e^{(n,S)}$ approaching zero.

In a composite channel, different channel states can be viewed as
virtual receivers, and therefore the expected capacity
is closely related to the capacity region of a broadcast channel
(BC).
In the broadcast system the channel from the input to the output of
receiver $s$ is
\[
P_{Y_s^n|X^n}(y_s^n|x^n) = P_{Y^n|X^n,S}(y_s^n|x^n,s).
\]
Under certain conditions, it is shown that the expected capacity of
a composite channel equals to the maximum weighted sum-rate over the
capacity region of the corresponding broadcast channel, where the
weight coefficient is the state probability $P(s)$
\cite[Theorem~1]{LiangJournal}. Using broadcast channel codes, the
expected capacity is derived in \cite{Shamai03} for a Gaussian
slow-fading channel and in \cite{Liang072} for a composite BSC.

The expected capacity is a meaningful metric if {\em partial}
received information is useful. For example, consider sending an
image using a multi-resolution (MR) source code over a composite
channel. Decoding all transmitted information leads to
reconstructions with the highest fidelity. However, in the case of
inferior channel quality, it still helps to decode partial
information and get a coarse reconstruction.

\subsection{End-to-End Distortion Metrics}
\label{sec:distortionMetric} Next we introduce alternative
end-to-end distortion metrics as performance measures for
transmission of a stationary ergodic source over a composite
channel. We denote by $\mathcal{V}$ the source alphabet and the
source symbols $\{ V^n = (V^{(n)}_1, V^{(n)}_2, \cdots, V^{(n)}_n)
\}_{n = 1}^{\infty}$ are generated according to a sequence of
finite-dimensional distributions $P(V^n)$, and then transmitted over
a composite channel $W^n: X^n \to (Y^n, S)$ with conditional output
distribution
\[
W^n(y^n,s|x^n) = P_S(s) P_{Y^n|X^n,S}(y^n|x^n,s).
\]
It is possible that the source generates symbols at a rate different
from the rate at which the channel transmits symbols, i.e. a
length-$n$ source sequence may be transmitted in $m$ channel uses
with $m\neq n$. The channel {\em bandwidth expansion ratio} is
defined to be $b=m/n$. For simplicity we assume $b=1$ in this and
the next two sections, but the discussions can be easily extended to
general cases with $b\neq 1$. The numerical examples in Section
\ref{sec:InterfaceBSC} will explicitly address this issue.

\subsubsection{Distortion versus Outage}
Here we design an encoder $f_n: V^n \to X^n$ that maps the source
sequence to the channel input. Note that the source and channel
encoders, whether joint or separate, do not have access to channel
state information $S$. However, the receiver can declare an outage
with probability $P^{(n)}_o$ based on CSIR. In non-outage states, we
design a decoder $\phi_n: (Y^n, S) \to \hat{V}^n$ that maps the
channel output to a source reconstruction. We say a distortion level
$D$ is outage-$q$ achievable if ${\displaystyle
\lim_{n\rightarrow\infty}P_o^{(n)}\leq q}$ and
\begin{equation}
\lim_{n \to \infty} \Pr \left\{ \left. (V^n, \hat{V}^n): d(V^n,
\hat{V}^n)
> D  \right|\textrm{no outage}\right\} = 0,
\label{eqn:outageDistortionConstraint}
\end{equation}
where $d(V^n, \hat{V}^n) = \frac{1}{n} \sum_{i=1}^n d(V_i,
\hat{V}_i)$ is the distortion measure between the source sequence
$V^n$ and its reconstruction $\hat{V}^n$. The {\em distortion versus
outage} $D_q$ is the infimum over all outage-$q$ achievable
distortions. In order to evaluate
\eqref{eqn:outageDistortionConstraint} we need the conditional
distribution $P(\hat{V}^n|V^n)$. Assuming the encoder $f_n$ and the
decoder $\phi_n$ are deterministic, this distribution is given by
\begin{equation}
\sum_{(X^n, Y^n, S)}  W^n(Y^n,S|X^n) \cdot \bs{1}\left\{X^n =
f_n(V^n),
\hat{V}^n = \phi_n(Y^n,S) \right\}
\label{eqn:conditionalDistribution}
\end{equation}
Here $\bs{1}\{\cdot\}$ is the indicator function. Note that the
channel statistics $W^n$ and the source statistics $P(V^n)$ are
fixed, so the code design is essentially the appropriate choice of
the outage states and the encoder-decoder pair $(f_n, \phi_n)$.

\subsubsection{Expected Distortion}
We denote by $D_S$ the achievable average distortion when the
channel is in state $S$, and it is given by
\begin{equation}
D_S = \lim_{n \to \infty} \sum P(V^n) W^n(Y^n|X^n,S) d(V^n,
\hat{V}^n), \label{eqn:defn_Ds}
\end{equation}
where the summation is over all ${(V^n, X^n, Y^n, \hat{V}^n)}$ such
that  $X^n = f_n(V^n)$ and $\hat{V}^n = \phi_n(Y^n,S)$. Notice that
the transmitter cannot access channel state information so the
encoder $f^n$ is independent of $S$; nevertheless the receiver can
choose different decoders $\phi_n(\cdot,S)$ based on CSIR.

In a composite channel, each channel state is assumed to be
stationary and ergodic, so for a fixed channel state $S$ we can
design source-channel codes such that $d(V^n, \hat{V}^n)$ approaches
a constant limit $D_S$ for large $n$; however, it is possible that
$d(V^n, \hat{V}^n)$ approaches different limits for different
channel states. The expected distortion metric captures the
distortion averaged over various channel states. Using the
conditional distribution $P(\hat{V}^n|V^n)$ in
\eqref{eqn:conditionalDistribution} and the definition of $D_S$ in
\eqref{eqn:defn_Ds}, the average distortion can be written
as\footnote{Assuming a bounded distortion measure, the exchange of
limit operation and expectation follows from the dominant
convergence theorem.}
\begin{equation}
\lim_{n \to \infty} \mathbb{E}_{(V^n, \hat{V}^n)} \left\{ d(V^n,
\hat{V}^n) \right\} = \sum_S P(S) D_S = \mathbb{E}_S D_S.
\label{eqn:expectedDistortion}
\end{equation}
The expected distortion $D^e$ is the infimum of all achievable
average distortions $\mathbb{E}_S D_S$.

\section{Source-Channel Separation and Interface: A New Perspective}
\label{sec:source-channel}

For transmission of a source over a channel, the system consists of
three concatenated blocks: the encoder $f_n$ that maps the source
sequence $V^n$ to the channel input $X^n$; the channel $W^n$ that
maps the channel input $X^n$ to channel output $Z^n$, and the
decoder $\phi_n$ that maps the channel output $Z^n$ to a
reconstruction of the source sequence $\hat{V}^n$. In contrast, a
separate source-channel coding scheme consists of five blocks. The
encoder $f_n$ is separated into a source encoder
\[
\tilde{f}_n: V^n \to \mathcal{M}_{n,t}= \{1, 2, \cdots, 2^{nR_t} \}
\]
and a channel encoder
\[
\hat{f}_n: \mathcal{M}_{n,t} = \{1, 2, \cdots, 2^{nR_t} \} \to X^n,
\]
where the index set $\mathcal{M}_{n,t}$ of size $2^{nR_t}$ serves as
both the source encoder output and the channel encoder input.
Equivalently, each index in $\mathcal{M}_{n,t}$ can be viewed as a
block of $nR_t$ bits \cite[Defn. 5]{LiangJournal}. The decoder
$\phi_n$ is also separated into a channel decoder $\hat{\phi}_n$ and
a source decoder $\tilde{\phi}_n$. The difference between a general
system and a separate source-channel coding system is summarized in
\refF{system3block}. \psfrag{Vn}{$V^n$} \psfrag{fn}{$f_n$}
\psfrag{Xn}{$X^n$} \psfrag{Zn}{$Z^n$} \psfrag{phin}{$\phi_n$}
\psfrag{Wn}{$W^n$} \psfrag{hatVn}{$\hat{V}^n$}
\psfrag{fntilde}{$\tilde{f}^n$} \psfrag{fnhat}{$\hat{f}^n$}
\psfrag{phintilde}{$\tilde{\phi}^n$}
\psfrag{phinhat}{$\hat{\phi}^n$}
\begin{figure}[htbp]
\begin{center}
\includegraphics[width=3.6in]{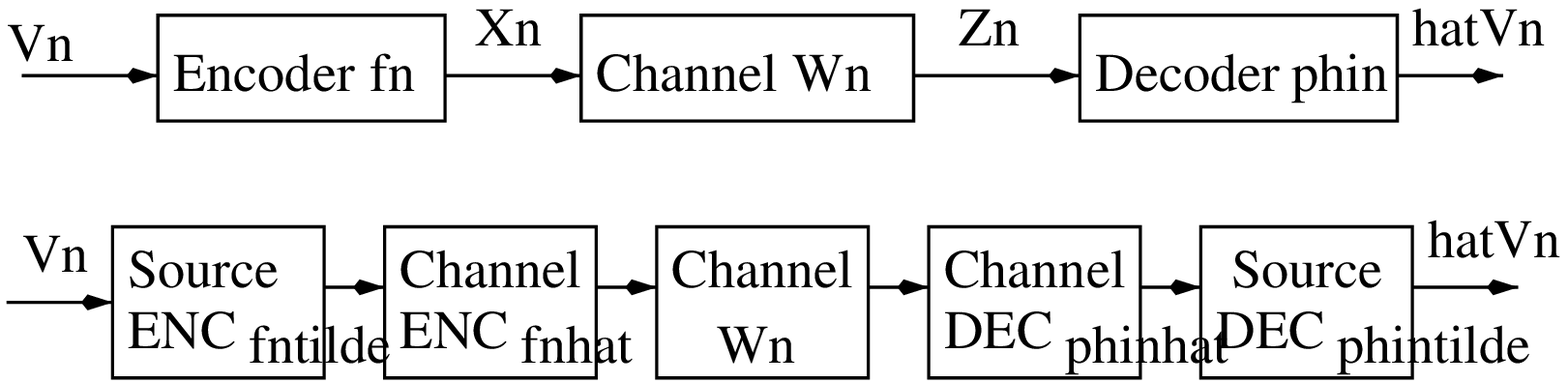}
\caption{Upper: general communication system with three blocks.
Lower: separate source-channel coding system with five blocks.}
\label{fig:system3block}
\end{center}
\end{figure}


Separation does not imply isolation - the source and channel
encoders and decoders still need to agree on certain aspects of
their respective designs. There are three interfaces through which
they exchange information, the negotiation interface, the
transmitter interface and the receiver interface. For classical
Shannon separation schemes with an end-to-end distortion target $D$,
these interfaces are summarized in Table
\ref{table:ShannonInterface}. The negotiation interface is a single
rate comparison between $R(D)$ and $C$. Since the Shannon capacity
definition requires that all transmitted information be correctly
decoded, the transmission rate $R_t$ is the same as the receiving
rate $R_r$. Assuming stationary and ergodic systems, these rates do
not depend on the blocklength $n$. However, these constraints can be
relaxed to include more source-channel transmission strategies as
separation schemes.
\begin{table}[htbp]
\caption{Interface for Shannon separation schemes} \centering
\begin{tabular}{l|p{2.6in}}
\hline Negotiation & source coding rate $R(D)$ and channel Shannon
capacity $C$ \\
\hline
Transmitter & $\mathcal{M}_{n,t} = \{1,2,\cdots, 2^{nR_t}\}$ \\
\hline
Receiver & $\mathcal{M}_{n,r} = \{1,2,\cdots, 2^{nR_r}\}$\\
\hline
\end{tabular}
\label{table:ShannonInterface}
\end{table}

In \cite{Vembu95} Vembu et al. proposed transmission schemes for
non-stationary source and channel models. The corresponding
interfaces are listed in Table \ref{table:VembuInterface}. Here the
negotiation interface is no longer a single number, but a sequence
of source and channel statistics for different blocklengths $n$. The
transmission and receiving rates are still the same, but now they
depend on the blocklength $n$.
\begin{table}[htbp]
\caption{Interface for Vembu separation schemes} \centering
\begin{tabular}{l|p{2.6in}}
\hline Negotiation & source entropy density $h_{V^n}(v^n)$ and
channel information density $i_{X^nW^n}(x^n; z^n)$ \\
\hline
Transmitter & $\mathcal{M}_{n,t} = \{1,2,\cdots, 2^{nc_n}\}$ \\
\hline
Receiver & $\mathcal{M}_{n,r} = \{1,2,\cdots, 2^{nc_n}\}$\\
\hline
\end{tabular}
\label{table:VembuInterface}
\end{table}

In Section \ref{sec:outageDistortion} we propose a separation scheme
for transmission of stationary ergodic sources over composite
channels, and prove its optimality under distortion versus outage
metrics. The interfaces of this scheme are shown in Table
\ref{table:OutageDistortionInterface}. The negotiation interface is
still a single number, but the channel should provide its capacity
versus outage-$q$ ($C_q$) \cite[Defn. 3]{LiangJournal} instead of
the Shannon capacity. The receiver interface includes an additional
outage indicator. In non-outage states, the channel decoder recovers
the channel input index with negligible error and delivers it to the
source decoder to achieve the end-to-end distortion target $D_q$. In
outage states the channel decoder shuts itself off and nothing
passes through the receiver interface.
\begin{table}[htbp]
\caption{Interface under distortion versus outage metric} \centering
\begin{tabular}{l|p{2.6in}}
\hline Negotiation & source coding rate $R(D_q)$ and
channel capacity versus outage-$q$ ($C_q$) \\
\hline
Transmitter & $\mathcal{M}_{n,t} = \{1,2,\cdots, 2^{nR}\}$ \\
\hline
Receiver & Outage indicator $I$. For non-outage states $\mathcal{M}_{n,r} = \mathcal{M}_{n,t}$\\
\hline
\end{tabular}
\label{table:OutageDistortionInterface}
\end{table}

In Section \ref{sec:InterfaceBSC} we study transmission of a binary
symmetric source over a composite BSC under the expected distortion
metric. One of the transmission schemes is to use a multi-resolution
source code and a broadcast channel code, with interfaces defined in
Table \ref{table:ExpectedDistortionInterface}. For the negotiation
interface, the channel provides the channel state probability $P(s)$
and the entire broadcast capacity region boundary. A point on the
boundary is a vector $(R_s)_{s \in \mathcal{S}}$ of achievable rates
in each channel state for a certain BC channel code. Based on the
distortion-rate function $D(R_s)$ of its multi-resolution code, the
source then chooses the rate vector $(R_s)$ to minimize the expected
distortion $\sum P(s)D(R_s)$. 
Without channel state information at the transmitter, the size of
the index set $\mathcal{M}_{n,t}$, i.e. the transmitter interface,
is fixed. Each index in $\mathcal{M}_{n,t}$ can be viewed as a block
of $nR_t$ bits. Different from the Shannon capacity definition, each
bit is only required to be successfully decoded by a subset of
channel states, not necessarily all states \cite[Defn.
5]{LiangJournal}. Consequently, the receiver can choose different
decoders based on CSIR, and the receiver interface
$\mathcal{M}_{n,s}$ depends on the channel state $s$.
\begin{table}[htbp]
\caption{Interface under expected distortion metric} \centering
\begin{tabular}{l|p{2.6in}}
\hline Negotiation & achievable distortion with multi-resolution
source code $D(R_s)$, broadcast channel capacity region $(R_s)_{s
\in \mathcal{S}}$ and corresponding channel state probability $P(s)$
\\
\hline
Transmitter & $\mathcal{M}_{n,t} = \{1,2,\cdots, 2^{nR_t}\}$ \\
\hline
Receiver & $\mathcal{M}_{n,s} = \{1,2,\cdots, 2^{nR_s}\}$ for channel state $s$\\
\hline
\end{tabular}
\label{table:ExpectedDistortionInterface}
\end{table}

Although the above schemes differ from each other in their choice of
interfaces, all of them retain the main advantage of separation -
modularity. For example, under the distortion versus outage metric,
there is a class of channels which can support rate $C_q$ with
probability no less than $(1-q)$. As long as $C_q$ exceeds the rate
distortion function $R(D_q)$, the source can be transmitted over any
channel within this class and be reconstructed at the destination
subject to the distortion versus outage constraint
\eqref{eqn:outageDistortionConstraint}. The source only need to know
$C_q$ to decide whether the constraint
\eqref{eqn:outageDistortionConstraint} can be satisfied, and the
source code design does not depend on any other channel statistics.
We can argue similarly for other transmission schemes. For all of
them, the encoder/decoder can be separated into a source
encoder/decoder and a channel encoder/decoder, as illustrated by the
five-block diagram in \refF{system3block}. A channel code can be
explicitly identified in this diagram, which includes the three
blocks in the middle. Note that the channel code might be designed
for generalized capacity definitions, not necessarily for the
Shannon capacity definition.

In contrast joint source-channel coding is a loose label that
encompasses all coding techniques where the source and channel
coders are not entirely separated. Consider the example of the {\em
direct transmission} of a complex circularly symmetric Gaussian
source, which we denote by $\mathcal{CN}(0, \sigma^2)$, over a
Gaussian channel with input power constraint $P$. The linear encoder
$X = f(V) = \sqrt{P/\sigma^2} V$ cannot be separated into a source
encoder and a channel encoder. Therefore this direct transmission is
an example of joint-source channel coding.

In Section \ref{sec:InterfaceBSC} we also propose two other schemes,
namely the {\em systematic coding} and {\em the quantization error
splitting}, for transmission of a binary symmetric source over a
composite BSC. These schemes are applicable because of the specific
system setup: the source alphabet is the same as the channel input
alphabet, and they do not apply if the BSC is replaced by some other
channels. We view them as joint source-channel coding schemes
because they lack flexibility and because we cannot identify a
three-block channel code as in previous examples. Nevertheless, the
interface concept can be extended to joint source-channel coding
schemes. The interface complexity, together with end-to-end
performance, provides two criterions to compare various schemes. We
defer the details to Section \ref{sec:InterfaceBSC}.

\section{Separation Optimality under Distortion versus Outage Metric}
\label{sec:outageDistortion} Consider transmission of a finite
alphabet stationary ergodic source $\{V_i\}_{i=1}^{\infty}$ over a
composite channel $\bs{W}$. In this section we show that the
classical Shannon separation theorem can be extended to
communication systems under the distortion versus outage metric.

\subsection{Lossless Transmission}
Denote by $C_q$ the channel capacity versus outage-$q$ and by
$H(\mathcal{V})$ the source entropy rate
\[
H(\mathcal{V}) = \lim_{n \to \infty} \frac1n H(V_1, V_2, \cdots,
V_n).
\]
We first consider the case of lossless transmission, i.e. $D=0$. The
distortion versus outage-$q$ constraint
\eqref{eqn:outageDistortionConstraint} now simplifies to
\begin{eqnarray*}
&& 
\Pr \left\{ \left. (V^n, \hat{V}^n): d(V^n,
\hat{V}^n)=0  \right|\textrm{no outage}\right\} \\
&=& 
\Pr \left\{ \left. V^n = \hat{V}^n \right|\textrm{no outage}\right\}
\to 1
\end{eqnarray*}
as $n$ approaches infinity.
\begin{theorem} \label{lemma:lossless_separation}
For lossless transmission, if $H(\mathcal{V}) < C_q$ then there
exists a sequence of blocklength-$n$ source-channel codes that
satisfy the outage-$q$ constraint
\begin{equation}
\lim_{n\rightarrow\infty}P_o^{(n)}\leq q, \,\,\, \lim_{n \to \infty}
\Pr \left\{ \left. V^n = \hat{V}^n \right|\textrm{{\em no
outage}}\right\} = 1; \label{eqn:lossless_q_outage}
\end{equation}
conversely, the existence of source-channel codes that satisfy the
above constraints also implies $H(\mathcal{V}) \le C_q$.
\end{theorem}

To prove the direct part, we construct a two-stage encoder $f_n$,
which involves a source encoder $\tilde{f}_n$ and a channel encoder
$\hat{f}_n$, and similarly for the decoder $\phi_n$.  The converse
of Theorem \ref{lemma:lossless_separation} then guarantees this
separate source-channel code essentially achieves optimal
performance, i.e. performance at least as good as any possible joint
coding scheme. The converse of the Shannon separation theorem
\cite[p. 217]{CoverIT} is established through Fano's inequality. It
is known that Fano's inequality fails to provide a tight lower bound
for error probability \cite{VerduH:94}, so here we use information
density to establish the converse for general channel models.
\\

\noindent \textbf{Proof}: In the following we denote $R =
H(\mathcal{V})$ and $C= C_q$ to simplify notation.

{\em Achievability}: Fix $\delta > 0$. Since the stationary ergodic
source satisfies asymptotic equipartition property (AEP) \cite[p.
51]{CoverIT}, for any $0 < \epsilon < 1$ and sufficiently large $n$,
there exists a source encoder
\[
\tilde{f}_n: V^n \to U \in \{1, 2, \cdots, 2^{n(R+\delta)} \}
\]
and a source decoder
\[
\tilde{\phi}_n: U \in \{1, 2, \cdots, 2^{n(R+\delta)} \} \to
\tilde{V}^n
\]
such that $\Pr\{ V^n \neq \tilde{V}^n \} \le \epsilon$. Here
$\tilde{V}^n$ is the decoder output of the stand-alone source code.
By definition of capacity versus outage \cite[Defn.3]{LiangJournal},
there exist channel codes with a channel encoder
\[
\hat{f}_n: U \in \{1, 2, \cdots, 2^{n(C-\delta)} \} \to X^n,
\]
outage indicator
\[
I: \mathcal{S}\to \{0,1\},
\]
and a channel decoder for non-outage states
\[
\hat{\phi}_n: Z^n = (Y^n,S) \to \hat{U} \in \{1, 2, \cdots,
2^{n(C-\delta)} \}
\]
such that for sufficiently large $n$, $P_o^{(n)} = \Pr \{I=0\} \le
q+\epsilon$ and $P_e^{(n)} = \Pr\{U\neq \hat{U}|I=1\} \le \epsilon$.
For sufficiently small $\delta$ we have $R+\delta<C-\delta$, which
guarantees the output of the source encoder $\tilde{f}_n$ always
lies in the domain of the channel encoder $\hat{f}_n$.

Now we concatenate the source encoder, channel encoder, channel
decoder and source decoder to form a communication system. We
declare an outage for the overall system whenever the channel is in
outage. For non-outage states, denote by $\hat{V}^n$ the source
reconstruction at the output of the overall system, given by
$\hat{V}^n = \tilde{\phi}_n \left( \hat{\phi}_n (Z^n) \right)$ with
$Z^n$ the channel output due to the channel input $X^n = \hat{f}_n
\left( \tilde{f}_n (V^n) \right)$. We have $P_o^{(n)}
 \le q+\epsilon$ and
\begin{eqnarray*}
&& \Pr \left\{ \left. V^n = \hat{V}^n \right|\textrm{no
outage}\right\} \\
&\geq& \Pr \left\{ \left. V^n = \hat{V}^n, U = \hat{U} \right|I =
1\right\} \\
&=& \Pr \left\{ \left. U = \hat{U} \right|I = 1\right\} \cdot \Pr
\left\{ \left. V^n = \hat{V}^n  \right| U = \hat{U}, I = 1\right\}
\\
&\geq& (1-\epsilon) (1-\epsilon).
\end{eqnarray*}
Since $\epsilon>0$ is arbitrary, \eqref{eqn:lossless_q_outage} is
proved.

{\em Converse}: Notice that
\[
\Pr \{ V^n = \hat{V}^n \} \ge  \left[ 1 - P_o^{(n)} \right] \cdot
\Pr \left\{ \left. V^n = \hat{V}^n \right|\textrm{no
outage}\right\},
\]
so the outage-$q$ constraint \eqref{eqn:lossless_q_outage} also
implies
\begin{equation}
\lim_{n \to \infty} \Pr \{ V^n = \hat{V}^n \} \ge  1-q.
\label{eqn:lossless_outage_constraint_noCSIR}
\end{equation}
The constraint \eqref{eqn:lossless_outage_constraint_noCSIR} is a
weaker condition than \eqref{eqn:lossless_q_outage} since it does
not require the outage event to be recognized by the decoder. In the
following we prove a stronger version of the converse: a
source-channel code with encoder $f_n$:~$V^n \to X^n$ and decoder
$\phi_n: Z^n = (Y^n, S) \to \hat{V}^n$ that satisfies the constraint
\eqref{eqn:lossless_outage_constraint_noCSIR} also implies
$H(\mathcal{V})\le C_q$, whether or not the outage event is
recognized.

Fix $\gamma > 0$. For any $0 < \epsilon < \gamma$, define the
typical set $A_\epsilon^{(n)}$ as
\begin{equation}
A_\epsilon^{(n)} = \left\{ v^n: \left| -\frac1n \log P_{V^n}(v^n) -
R \right| < \epsilon \right\}. \label{eqn:TypicalSetDefinition}
\end{equation}
For any $v^n \in \mathcal{V}^n$, define
\[
D(v^n) = \left\{ Z^n \in \mathcal{Z}^n: \phi_n(z^n) = v^n \right\}
\]
as the decoding region for $v^n$ and
\begin{equation}
B(v^n) = \left\{ Z^n \in \mathcal{Z}^n: \frac1n i_{X^nW^n} \left(
f_n(v^n); z^n \right) \le R - 2\gamma \right\}.
\label{eqn:BvnDefinition}
\end{equation}
Then we have
\begin{eqnarray}
&& \Pr \left\{ \frac1n i_{X^nW^n} (X^n; Z^n) \le R - 2\gamma
\right\}
\nonumber \\
&=& \sum_{(v^n, z^n)} P_{V^n}(v^n) W^n(z^n|f_n(v^n)) \cdot \bs{1}
\left\{z^n \in B(v^n) \right\} \nonumber \\
&=& \left( \sum_{\Gamma_1} + \sum_{\Gamma_2} + \sum_{\Gamma_3}
\right) P_{V^n}(v^n) W^n (z^n|f_n(v^n)), \label{eqn:threeParts}
\end{eqnarray}
where $\bs{1}\{\cdot\}$ is the indicator function. In
\eqref{eqn:threeParts} we divide the summation into three regions
\begin{eqnarray*}
\Gamma_1&=& \left\{(v^n, z^n):  v^n \notin A_\epsilon^{(n)}, z^n \in
B(v^n) \right\}, \\
\Gamma_2&=& \left\{(v^n, z^n): v^n \in A_\epsilon^{(n)}, z^n \in
B(v^n) \cap D(v^n) \right\}, \\
\Gamma_3&=& \left\{(v^n, z^n): v^n \in A_\epsilon^{(n)}, z^n \in
B(v^n) \cap D^c(v^n) \right\},
\end{eqnarray*}
where $D^c(v^n)$ is the complement of the decoding region $D(v^n)$.
We can bound the summation over each region as follows. For the
first term, we have
\begin{equation}
\sum_{\Gamma_1} P_{V^n}(v^n) W^n (z^n|f_n(v^n)) \le 1 - P_{V^n}
\left\{ A_\epsilon^{(n)} \right\} \le \epsilon \label{eqn:part1}
\end{equation}
for sufficiently large $n$ as a result of AEP \cite[p.52]{CoverIT}.
For the second term, we have
\begin{eqnarray}
P_{V^n}(v^n) &\le& 2^{-n (R-\epsilon)} \le 2^{-n (R-\gamma)}
\label{eqn:PvnUpperBound}
\\
W^n(z^n|f_n(v^n)) &\le& 2^{n(R-2\gamma)} P_{Z^n} (z^n)
\label{eqn:WznUpperBound}
\end{eqnarray}
for any $(v^n, z^n) \in \Gamma_2$, where \eqref{eqn:PvnUpperBound}
is a property of the typical set $A_\epsilon^{(n)}$
\eqref{eqn:TypicalSetDefinition}, and \eqref{eqn:WznUpperBound} is
obtained from \eqref{eqn:BvnDefinition} and the information density
definition \eqref{eqn:infoden}. The decoding regions of different
$v^n$ do not overlap, and therefore
\begin{equation}
\sum_{\Gamma_2} P_{V^n}(v^n) W^n (z^n|f_n(v^n)) \le \sum_{\Gamma_2}
2^{-n\gamma} P_{Z^n}(z^n) \le 2^{-n\gamma}. \label{eqn:part2}
\end{equation}
For the third term,
\begin{eqnarray}
&& \sum_{\Gamma_3} P_{V^n}(v^n) W^n (z^n|f_n(v^n)) \nonumber \\
&\le& \sum_{v^n}
P_{V^n}(v^n) W^n (D^c(v^n)|f_n(v^n)) \nonumber \\
&=& \Pr\{V^n \neq \hat{V}^n\}. \label{eqn:part3}
\end{eqnarray}
Combining \eqref{eqn:threeParts}-\eqref{eqn:part1},
\eqref{eqn:part2}-\eqref{eqn:part3}, we obtain
\[
\Pr\{V^n \neq \hat{V}^n\} \ge \Pr \left\{ \frac1n i_{X^nW^n} (X^n;
Z^n) \le R - 2\gamma \right\} - 2^{-n\gamma} - \epsilon.
\]
Let $\epsilon \to 0$ and $n \to \infty$, since the constraint
\eqref{eqn:lossless_outage_constraint_noCSIR} requires the error
probability of the source-channel code to be upper bounded by $q$,
we conclude
\[
\lim_{n \to \infty} \Pr \left\{ \frac1n i_{X^nW^n} (X^n; Z^n) \le R
- 2\gamma \right\} \le q.
\]
Since $\gamma > 0$ is arbitrary, by definition of $C_q$ we must have
$H(\mathcal{V}) = R \le C_q$. \qed

\subsection{Lossy Transmission}
For the case of lossy transmission $(D>0)$, we focus on discrete
memoryless sources (DMS) $\{V_i\}_{i=1}^{\infty}$ and recall the
definition of a source rate-distortion function as \cite[p.
342]{CoverIT}
\begin{equation}
R(D) = \min_{P(\hat{V}|V): \mathbb{E}d(V,\hat{V}) \le D}
I(V;\hat{V}). \label{eqn:sourceDistortion}
\end{equation}
Extensions to sources with memory follow the procedures in
\cite[Sec. 7.2]{Berger}. Occasionally we also use the notation
$R(V,D)$ to specify the source distribution.
For discrete memoryless source and channel models, it is shown that
if $R(D) < C$ then the source can be transmitted over the channel
subject to an {\em average fidelity criterion}
\begin{equation}
\mathbb{E}\left\{d(V^n, \hat{V}^n)\right\} \le D.
\label{eqn:avgFidelity}
\end{equation}
Conversely, if the transmission satisfies the average fidelity
criterion, we also conclude $R(D) \le C$ \cite[p. 130]{CsiszarK:81}.
Next we consider composite channel models and generalized distortion
metrics.

\begin{theorem} \label{lemma:lossy_separation}
Denote by $R(D_q)$ the rate-distortion function
\eqref{eqn:sourceDistortion} of a discrete i.i.d. source evaluated
at distortion level $D_q$. If $R(D_q) < C_q$ the source can be
transmitted over a composite channel subject to the outage
constraint \eqref{eqn:outageDistortionConstraint}
\begin{eqnarray*}
&& \lim_{n\rightarrow\infty} P_o^{(n)}\leq q, \\
&& \lim_{n \to \infty} \Pr \left\{ \left. (V^n, \hat{V}^n): d(V^n,
\hat{V}^n) > D_q  \right|\textrm{{\em no outage}}\right\} = 0;
\end{eqnarray*}
conversely, the existence of source-channel codes that satisfy the
above constraints also implies $R(D_q) \le C_q$.
\end{theorem}


The proof of the direct part of Theorem \ref{lemma:lossy_separation}
is similar to that of Theorem \ref{lemma:lossless_separation}. The
new element is a change from lossless source coding to lossy source
coding. In the rate distortion theory for source coding, one often
imposes the {\em average fidelity criterion}
$\mathbb{E}\left\{d(V^n, \tilde{V}^n)\right\} \le D$,
where $\tilde{V}^n$ is the source reconstruction sequence. The main
challenge here is to satisfy the condition
\eqref{eqn:outageDistortionConstraint} which is based on the tail of
the distortion distribution rather than on its mean. So for source
coding, instead of the global average fidelity criterion
\eqref{eqn:avgFidelity}, we impose the following local
$\epsilon$-{\em fidelity criterion} \cite[p. 123]{CsiszarK:81}
\begin{equation}
\Pr\left\{(V^n, \tilde{V}^n): d(V^n, \tilde{V}^n) \le D\right\} \ge
1 - \epsilon. \label{eqn:epsilonfidelity}
\end{equation}
%
%
It is well known that
for any $\delta>0$ there exist source codes with rate
$R<R(D)+\delta$ which satisfy the average fidelity criterion
\eqref{eqn:avgFidelity} \cite[p. 351]{Cover72}. To prove the direct
part of Theorem \ref{lemma:lossy_separation}, we need a stronger
result \cite[p. 125]{CsiszarK:81}:
%
%
for any $0 <
\epsilon < 1$ and $\delta > 0$, there exists source encoder
\[
\tilde{f}_n: V^n \to U \in \left\{1, 2, \cdots, 2^{n[R(D)+\delta]}
\right\}
\]
and source decoder
\[
\tilde{\phi}_n: U \in \left\{1, 2, \cdots, 2^{n[R(D)+\delta]}
\right\} \to \tilde{V}^n
\]
such that $\Pr\left\{d(V^n, \tilde{V}^n) \le D\right\} \ge 1 -
\epsilon$. We can then construct channel codes for capacity versus
outage-$q$ and concatenate it with the $\epsilon$-fidelity source
code to satisfy the outage constraint
\eqref{eqn:outageDistortionConstraint}, similarly as in Theorem
\ref{lemma:lossless_separation}. \\

Next we consider the converse of Theorem
\ref{lemma:lossy_separation}. Similar to the case of lossless
transmission, we prove a stronger version of the converse which does
not require outage events to be recognized by the decoder. Notice
that
\begin{eqnarray*}
&& \Pr \left\{ (V^n, \hat{V}^n): d(V^n, \hat{V}^n) \le D \right\} \\
&\ge& \left[ 1 - P_o^{(n)} \right] \cdot \Pr \left\{ \left. d(V^n,
\hat{V}^n) \le D \right|\textrm{no outage}\right\},
\end{eqnarray*}
so the outage constraint \eqref{eqn:outageDistortionConstraint}
implies
\begin{equation}
\lim_{n \to \infty} \Pr \left\{ (V^n, \hat{V}^n): d(V^n, \hat{V}^n)
\le D \right\}  \ge  1-q. \label{eqn:lossy_outage_constraint_noCSIR}
\end{equation}
We show the constraint \eqref{eqn:lossy_outage_constraint_noCSIR}
also implies $R(D_q) \le C_q$.

A brief review of the converse of the Shannon separation theorem
\cite[p.130]{CsiszarK:81} helps to highlight the new challenges
here. For transmission of a DMS over a DMC under the average
fidelity criterion \eqref{eqn:avgFidelity}, the converse is
established through the following chain of inequalities
\begin{eqnarray}
C &\ge& \frac{1}{n} I(X^n; Z^n)  \label{eqn:capacityInequality}\\
&\ge& \frac{1}{n} I(V^n; \hat{V}^n) \label{eqn:InfoProcessInequality} \\
&\ge& R(D) \label{eqn:rateDistortionInequality},
\end{eqnarray}
where \eqref{eqn:capacityInequality} is a result of \cite[Lemma
8.9.2]{CoverIT}, \eqref{eqn:InfoProcessInequality} is from the
Markov-chain relationship $V^n \to X^n \to Z^n \to \hat{V}^n$ and
the data processing inequality \cite[Theorem 2.8.1]{CoverIT}, and
\eqref{eqn:rateDistortionInequality} is from the convexity of a
rate-distortion function \cite[p.350]{CoverIT}.

We face two problems when trying to extend the previous approach to
composite channel models. First the capacity versus outage-$q$ is
defined through information density instead of mutual information,
and the data processing inequality does not have a counterpart in
terms of information density. Hence we need to refine the lower
bound of error probability in terms of information density following
a similar approach in the lossless case.

Second the rate distortion function \eqref{eqn:sourceDistortion} is
defined through an average fidelity criterion but the source and its
reconstruction satisfy the $q$-fidelity criterion
\eqref{eqn:lossy_outage_constraint_noCSIR}. In this regard we
consider the {\em joint type} \cite[p.~279]{CoverIT} or empirical
probability distribution $\tilde{P} (V_*, \hat{V}_*)$ induced by a
pair of sequences $(v^n, \hat{v}^n)$, where $v^n$ is a {\em strong
typical sequence} \cite[p. 33]{CsiszarK:81} and $\hat{v}^n$ is the
reconstruction sequence satisfying $d(v^n, \hat{v}^n) \le D$.
Briefly speaking, by definition of joint type the distribution
$\tilde{P}$ satisfies the average fidelity criterion $\mathbb{E}
d(V_*, \hat{V}_* ) \le D$. By definition of strong typicality the
marginal distribution $\tilde{P} \left(V_* \right)$ is ``close" to
the true source distribution $P(V)$, so the corresponding
rate-distortion functions $R(V_*, D)$ and $R(V,D)$ are also ``close"
to each other by continuity. This idea is formalized in the next
proof, prior to which we must define the notion of a {\em strong
typical sequence}:

\begin{definition} {\em \cite[p. 33]{CsiszarK:81} For a
random variable $V$ with alphabet $\mathcal{V}$ and distribution
$p(v)$, a sequence $v^n \in \mathcal{V}^n$ is said to be
$\delta$-strongly typical if
\begin{itemize}
\item for all $a \in \mathcal{V}$ with $p(a)>0$,
\[
\left| \frac{1}{n}N(a|v^n) - p(a) \right| < \delta;
\]
\item for all $a \in \mathcal{V}$ with $p(a)=0$, $N(a|v^n)=0$.
\end{itemize}
$N(a|v^n)$ is the number of occurrences of the symbol $a$ in $v^n$.
}
\end{definition}
The set of such sequences will be denoted by $T^n_{[V]_\delta}$, or
$T^n_{\delta}(V),$ or simply $T^n_{[V]}$. Let $v_i$, $1 \le i \le
n$, be drawn i.i.d. according to $p(v)$. Following the strong law of
large numbers, it is seen that for any $\epsilon > 0$, $\delta > 0$
and sufficiently large $n$, we have
\[
P_{V^n} \left( T^n_{[V]_\delta} \right) \ge 1 - \epsilon.
\]
By definition of strong typicality,
%
for any sequence $v^n \in T^n_{[V]_\delta}$
we also have
\begin{equation}
P_{V^n}(v^n) \le 2^{-n[H(V)-\delta']}, \label{eqn:pvn_bound}
\end{equation}
where
\[
\delta' = - \delta \sum_{a: \, p(a)>0} \log p(a) > 0.
\]
The upper bound \eqref{eqn:pvn_bound} is an immediate result by
noticing that
\[
\log P_{V^n}(v^n) = \sum_{a: \, p(a)>0} N(a|v^n) \log p(a)
\]
and $v^n \in T^n_{[V]_\delta}$ implies $N(a|v^n) > n \left[ p(a) -
\delta \right]$.\\

The definition of a strong typical sequence can be extended to
jointly distributed variables.
\begin{definition} {\em \cite[p.359]{CoverIT} A pair of sequences $(v^n, \hat{v}^n) \in
\mathcal{V}^n \times \hat{\mathcal{V}}^n$ is said to be
$\delta$-strongly typical with respect to the distribution $p(v,
\hat{v})$ on $\mathcal{V}\times \hat{\mathcal{V}}$ if
\begin{itemize}
\item for all $(a,b) \in \mathcal{V}\times \hat{\mathcal{V}}$ with $p(a,b)>0$ we have
\[
\left| \frac{1}{n}N(a,b|v^n, \hat{v}^n) - p(a,b) \right| <
\delta
\]
\item for all $(a,b) \in \mathcal{V}\times \hat{\mathcal{V}}$ with $p(a,b)=0$, $N(a,b|v^n, \hat{v}^n)=0$.
\end{itemize}
$N(a, b|v^n, \hat{v}^n)$ is the number of occurrences of the pair
$(a, b)$ in the pair of sequences $(v^n, \hat{v}^n)$. }
\end{definition}
The set of such sequences will be denoted by
$T^n_{[V,\hat{V}]_\delta}$, or $T^n_{\delta}(V, \hat{V})$, or
$T^n_{\delta}$ if the variables are clear from
context.\\

\noindent \textbf{Proof of Theorem \ref{lemma:lossy_separation}}: In
the following we denote $R = R(D_q)$, $D= D_q$ and $C= C_q$ to
simplify notation.

{\em Converse}: Consider a source-channel code with encoder
$f_n$:~$V^n \to X^n$ and decoder $\phi_n$: $Z^n = (Y^n, S) \to
\hat{V}^n$ that satisfy the outage constraint
\eqref{eqn:lossy_outage_constraint_noCSIR}. We assume both the
encoder and the decoder are deterministic.

Fix $\gamma > 0$. Consider $0 < \epsilon < (\gamma/4)$ and
\[
0 < \delta < - \frac{\epsilon}{\sum_{a: p(a)>0} \log p(a)}.
\]
From \eqref{eqn:pvn_bound}, for any $v^n \in T^n_{[V]_\delta}$ the
choice of $\delta$ ensures
\[
P_{V^n}(v^n) \le 2^{-n[H(V)-\epsilon]}.
\]
For each $v^n \in \mathcal{V}^n$, define
\[
D(v^n) = \{ z^n \in \mathcal{Z}^n: d(v^n, \phi_n(z^n)) \le D \}
\]
as the set of channel outputs which are mapped to {\em valid} source
reconstructions, i.e. those within distortion $D$ of the original
source sequence $v^n$. We also define
\[
B(v^n) = \left\{ z^n \in \mathcal{Z}^n: \frac{1}{n}
i_{X^nW^n}(f_n(v^n); z^n) \le R - 2\gamma \right\}.
\]
Next we derive an upper bound on the probability of {\em valid}
pairs of sequences. We have
\begin{eqnarray}
&& \Pr \left\{ d(V^n, \hat{V}^n) \le D \right\} \nonumber \\
&=& \sum_{(v^n, z^n)} P_{V^n}(v^n) W^n(z^n|f_n(v^n)) \cdot \bs{1}
\left\{z^n \in D(v^n) \right\} \nonumber \\
&=& \left( \sum_{\Gamma_1} + \sum_{\Gamma_2} + \sum_{\Gamma_3}
\right) P_{V^n}(v^n) W^n (z^n|f_n(v^n)),
\label{eqn:lossy_threeParts}
\end{eqnarray}
In \eqref{eqn:lossy_threeParts} we divide the summation into three
regions
\begin{eqnarray*}
\Gamma_1&=& \left\{(v^n, z^n):  v^n \notin T^n_{[V]_\delta}, z^n \in
D(v^n) \right\}, \\
\Gamma_2&=& \left\{(v^n, z^n): v^n \in T^n_{[V]_\delta}, z^n \in
B(v^n) \cap D(v^n) \right\}, \\
\Gamma_3&=& \left\{(v^n, z^n): v^n \in T^n_{[V]_\delta}, z^n \in
B^c(v^n) \cap D(v^n) \right\},
\end{eqnarray*}
where $B^c(v^n)$ is the complement of the region $B(v^n)$. We can
bound the summation over each region as follows. For sufficiently
large $n$,  the first term is bounded by
\begin{equation}
\sum_{\Gamma_1} P_{V^n}(v^n) W^n (z^n|f_n(v^n)) \le 1 - P_{V^n}
\left( T^n_{[V]_\delta} \right) \le \epsilon.
\label{eqn:lossy_part1}
\end{equation}
In the second term, for any $(v^n, z^n) \in \Gamma_2$ we have
\begin{eqnarray*}
P_{V^n}(v^n) &\le& 2^{-n [H(V)-\epsilon]} \\
W^n(z^n|f_n(v^n)) &\le& 2^{n(R-2\gamma)} P_{Z^n} (z^n),
\end{eqnarray*}
therefore
\begin{eqnarray}
&& \sum_{\Gamma_2} P_{V^n}(v^n) W^n (z^n|f_n(v^n))\nonumber \\
&\le& 2^{-n[H(V)-\epsilon-R+2\gamma]} \sum_{\Gamma_2} P_{Z^n}(z^n).
\label{eqn:repeatcounting}
\end{eqnarray}
Notice that, in contrast to the lossless case, the regions $D(v^n)$
are not necessarily disjoint; hence the summation in
\eqref{eqn:repeatcounting} may count the same sequence $z^n$ more
than once for every $v^n \in T^n_{[V]_\delta}$ satisfying $d(v^n,
\phi_n(z^n)) \le D$. In the following we give an upper bound of this
repeated counting.

For any $(v^n, z^n) \in \Gamma_2$ and the corresponding decoder
output $\hat{v}^n = \phi_n(z^n)$, we define a pair of random
variables $(\tilde{V}, \hat{\tilde{V}})$ with joint distribution
\[
\tilde{P}(a, b) = P_{v^n, \hat{v}^n}(a,b) = N(a, b|v^n,
\hat{v}^n)/n,
\]
where $N(a, b|v^n, \hat{v}^n)$ is the number of occurrences of the
pair $(a, b)$ in the pair of sequences $(v^n, \hat{v}^n)$.
$\tilde{P}$ is also called the {\em joint type} or empirical
probability distribution of $(v^n, \hat{v}^n)$
\cite[p.~279]{CoverIT}. Since for every $(a,b) \in \mathcal{V}
\times \hat{\mathcal{V}}$, there are at most $(n+1)$ possible values
$\{0,1,\cdots,n\}$ for $N(a, b|v^n, \hat{v}^n)$, the number of
different types is upper bounded by
$(n+1)^{|\mathcal{V}|\cdot|\hat{\mathcal{V}}|}$.

For every fixed $\hat{v}^n$, the number of sequences $v^n \in
\mathcal{V}^n$ with joint type $\tilde{P}$ is upper bounded by
$2^{nH(\tilde{V}|\hat{\tilde{V}})}$ \cite[Lemma~1.2.5]{CsiszarK:81}.
When ranging over $(v^n, z^n) \in \Gamma_2$, we can choose the pair
of sequences $(v^n_*, z^n_*)$, the corresponding decoder output
$\hat{v}^n_*$ and the pair of induced random variables $(V_*,
\hat{V}_*)$ that maximizes $H(\tilde{V}|\hat{\tilde{V}})$. So the
repeated counting for each fixed $z^n$ is upper bounded by
\[
(n+1)^{|\mathcal{V}|\cdot|\hat{\mathcal{V}}|}
2^{n[H(V_*|\hat{V}_*)]}
\]
%
%
and we continue \eqref{eqn:repeatcounting} to obtain
\begin{eqnarray}
&& \sum_{\Gamma_2} P_{V^n}(v^n) W^n (z^n|f_n(v^n))\nonumber \\
&\le& (n+1)^{|\mathcal{V}|\cdot|\hat{\mathcal{V}}|} \cdot
2^{-n[H(V)-\epsilon-R+2\gamma-H(V_*|\hat{V}_*)]} \sum_{z^n}
P_{Z^n}(z^n) \nonumber \\
& \le& (n+1)^{|\mathcal{V}|\cdot|\hat{\mathcal{V}}|} \cdot
2^{-n[H(V)-H(V_*)+I(V_*;\hat{V}_*)-R+2\gamma-\epsilon]}.
\label{eqn:term2_second}
\end{eqnarray}
For sufficiently large $n$ we have
\begin{equation}
(n+1)^{|\mathcal{V}|\cdot|\hat{\mathcal{V}}|} \le 2^{n\epsilon}.
\label{eqn:lossy_bound1}
\end{equation}
Obviously $v^n_* \in T^n_{[V]_\delta}$, so for any letter $a$ in the
alphabet $\mathcal{V}$ we have $|P_{V_*}(a) - p(a)|< \delta$. By
continuity of the entropy function,
\begin{equation}
|H(V)-H(V_*)| < \epsilon \label{eqn:lossy_bound2}
\end{equation}
for sufficiently small $\delta$. Since $\mathbb{E}d(V_*, \hat{V}_*)
= d(v^n_*, \hat{v}^n_*) \le D$, by definition of rate-distortion
function $I(V_*;\hat{V}_*) \ge R(V_*, D)$, where the notation
$R(V_*, D)$ emphasizes the source distribution is $P_{V_*}$.
Furthermore we know the rate-distortion function is continuous with
respect to the source distribution \cite[p. 124]{CsiszarK:81}, for
sufficiently small $\delta$
\begin{equation}
R = R(V, D) < R(V_*, D) + \epsilon \le I(V_*;\hat{V}_*) + \epsilon.
\label{eqn:lossy_bound3}
\end{equation}
Combine \eqref{eqn:term2_second}-\eqref{eqn:lossy_bound3} and notice
that $0 < \epsilon < (\gamma/4)$, we obtain
\begin{equation}
\sum_{\Gamma_2} P_{V^n}(v^n) W^n (z^n|f_n(v^n)) \le 2^{-n\gamma}.
\label{eqn:lossy_part2}
\end{equation}
For the third term,
\begin{eqnarray}
&& \sum_{\Gamma_3} P_{V^n}(v^n) W^n (z^n|f_n(v^n)) \nonumber \\
&\le& \sum_{v^n}
P_{V^n}(v^n) W^n (B^c(v^n)|f_n(v^n)) \nonumber \\
&=& 1 - \Pr \left\{\frac{1}{n} i_{X^nW^n} (X^n; Z^n) \le R - 2\gamma
\right\}. \label{eqn:lossy_part3}
\end{eqnarray}
Since the source-channel code satisfies the outage distortion
constraint \eqref{eqn:lossy_outage_constraint_noCSIR}, from
\eqref{eqn:lossy_part1}, \eqref{eqn:lossy_part2} and
\eqref{eqn:lossy_part3}, for sufficiently large $n$
\begin{eqnarray*}
&& 1 - q - \epsilon \\
&\le&
\Pr \left\{ d(V^n, \hat{V}^n) \le D \right\} \\
&\le& \epsilon + 2^{-n\gamma} + 1 - \Pr \left\{\frac{1}{n}
i_{X^nW^n} (X^n; Z^n) \le R - 2\gamma \right\}.
\end{eqnarray*}
Let $\epsilon \to 0$ and $n \to \infty$, we conclude
\[
\lim_{n \to \infty} \Pr \left\{ \frac1n i_{X^nW^n} (X^n; Z^n) \le R
- 2\gamma \right\} \le q,
\]
which, by definition of $C_q$, implies $R = R(D_q) \le C_q$. \qed


Note that although Theorem \ref{lemma:lossy_separation} is derived
for sources with finite alphabets and bounded distortion measures,
the result can be generalized to continuous-alphabet sources and
unbounded distortion measures using the technique of \cite[Ch.
7]{Gallagerbook}.

For our strategy the outage states are recognized by the receiver,
which can request a retransmission or simply reconstruct the source
symbol by its mean -- hence the distortion is the variance of the
source symbol. If we concatenate the source code in the direct part
of Theorem \ref{lemma:lossless_separation} and
\ref{lemma:lossy_separation} with a channel code based on
$\epsilon$-capacity \cite{VerduH:94}, the relaxed constraints
\eqref{eqn:lossless_outage_constraint_noCSIR} and
\eqref{eqn:lossy_outage_constraint_noCSIR} can still be satisfied.
However, there is a subtle difference. The receiver cannot recognize
the outage events in the latter strategy and the reconstruction
based on the decoded symbols, possibly in error, may lead to large
distortions.

\subsection{Example: Transmission of a Gaussian Source over a Slowly
Fading Gaussian Channels}
\label{sec:ExampleGaussianSourceSlowFadingChannel}
\psfrag{Vn}{$V^n$} \psfrag{fn}{$f_n$} \psfrag{CN}{$\mathcal{CN}(0,
\sigma^2)$} \psfrag{Xn}{$X^n$} \psfrag{Yn}{$Y^n$}
\psfrag{phin}{$\phi_n$} \psfrag{pga}{$p(\gamma)$}
\psfrag{hatVn}{$\hat{V}^n$}
\begin{figure}[htbp]
\begin{center}
\includegraphics
[width = 4.5in] {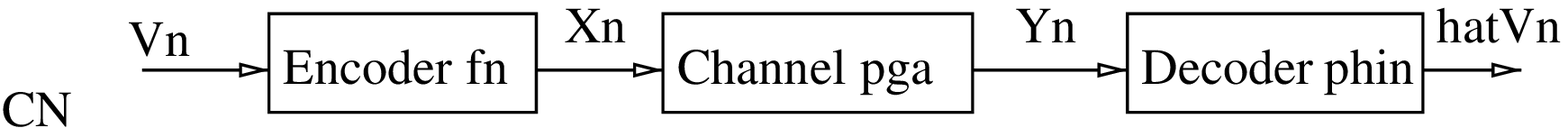} \caption{Transmission of
Gaussian source over slow-fading Gaussian channels}
\label{fig:slowFading}
\end{center}
\end{figure}

\subsubsection{Distortion verus Outage Metric}
We illustrate the separate source and channel codes
constructed in Theorem \ref{lemma:lossy_separation} by the following
example. As shown in \refF{slowFading}, a Gaussian source
$\mathcal{CN}(0, \sigma^2)$ is transmitted over a Rayleigh
slow-fading Gaussian channel with fading distribution $p(\gamma) =
\left(1/\bar{\gamma}\right) e^{-\gamma/\bar{\gamma}}$, where
$\bar{\gamma}$ is the average channel power gain. The transmitter
has a power constraint $P$. The additive Gaussian noise is i.i.d.
and normalized to have unit variance. The channel realization is
only known to the receiver but not the transmitter. In this example
we index each channel by the power gain $\gamma$, which has the same
role as the previous channel index $s$. We consider the case where
the source block length is the same as the channel block length,
i.e. the {\em bandwidth expansion ratio} $b$ equals to $1$.

For an outage probability $q$ the corresponding threshold of channel
gain is $\gamma_q = - \bar{\gamma} \log(1-q)$, so in non-outage
states the channel can support a rate of
\begin{equation} \label{eqn:Rq}
C_q = \log(1+P\gamma_q) = \log \left[ 1 - P \bar{\gamma} \log (1-q)
\right].
\end{equation}
The rate distortion function of a complex Gaussian source is given
by $R(D_q) = \log (\sigma^2/D_q)$. From Theorem
\ref{lemma:lossy_separation} if
\begin{equation}
\sigma^2/D_q < 1 - P \bar{\gamma} \log (1-q),
\label{eqn:requirement}
\end{equation}
then the outage constraint \eqref{eqn:outageDistortionConstraint}
can be satisfied by concatenation of a source code at rate $R(D_q)$
and a channel code at rate $C_q$. 

It is well known that the uncoded scheme is optimal for transmission
of a Gaussian source over a Gaussian channel when the bandwidth
expansion ratio $b=1$ \cite{Mittal02, Gastpar03}. The optimality is
in the sense that a linear code $X = \sqrt{P/\sigma^2}V$ can achieve
the minimum distortion
\begin{equation}
D_\gamma^* = \frac{\sigma^2}{1+P\gamma}
\label{eqn:optimalDistortion}
\end{equation}
for each channel state $\gamma$. It is easily seen that the optimal
uncoded scheme also requires \eqref{eqn:requirement} to satisfy the
outage distortion constraint. In summary, a separate source-channel
coding scheme meets the outage constraint
\eqref{eqn:outageDistortionConstraint} if $R(D_q) < C_q$;
if $R(D_q) > C_q$ then the constraint can never be satisfied even
for optimal joint source-channel coding. The result can be extended
to slow-fading Gaussian channels with any fading distribution
$p(\gamma)$.\\

\subsubsection{Expected Distortion Metric} Unlike the distortion versus outage metric,
source-channel separation does not hold for the expected distortion
metric. In the following we analyze the expected distortion of
optimal uncoded schemes and separate source-channel coding schemes.

{\em Optimal joint source-channel coding}: The uncoded scheme with a
direct mapping $X = \sqrt{P/\sigma^2}V$ can achieve the minimum
distortion \eqref{eqn:optimalDistortion} for each channel state
$\gamma$, and hence the optimal expected distortion
\begin{equation}
(D^e)^* = \int_0^\infty \frac{\sigma^2 e^{-\gamma/\bar{\gamma}}}
{1+P\gamma} \cdot
 \frac{d\gamma}{\bar{\gamma}}
= \frac{\sigma^2 e^{1/P\bar{\gamma}} }{P\bar{\gamma}}
\text{Ei}\left( \frac{1}{P\bar{\gamma}} \right),
\label{eqn:OptimalDe}
\end{equation}
with $\text{Ei}(x) = \int_x^\infty \left( \frac{e^{-t}}{t} \right)
dt$ the {\em exponential integral function}.\\

{\em Separation scheme with channel code for capacity versus
outage}: Consider using a channel code at rate $C_q$ for capacity
versus outage and a source code at the same rate. 
With probability $q$ the channel is in outage so the receiver
estimates the transmitted source symbols by its mean to achieve a
distortion of $\sigma^2$. With probability $(1-q)$ the channel can
support the rate $C_q$ and the end-to-end distortion is $D_q =
D(C_q)$. The overall expected distortion is averaged over the
non-outage and outage states, i.e. $D^e_1(q) = q\sigma^2 + (1-q)
D_q$.

The minimum achievable distortion of this strategy is obtained by
optimizing $D^e_1(q)$ over $q \in (0,1)$, i.e.
\begin{equation}
D^e_1 =  \min_{0<q<1} D^e_1(q) = \min_{0<q<1} q \sigma^2 +
\frac{(1-q) \sigma^2} {1 - P \bar{\gamma} \log(1-q)}.
\label{eqn:outageDistOpt}
\end{equation}
The solution is to use a channel code with outage probability
\begin{equation}
q^*_D = 1 - \exp \left\{ - \frac{2}{1+\sqrt{1+4P\bar{\gamma}}}
\label{eqn:qDstar} \right\}.
\end{equation}
One might be tempted to think that the channel should optimize its
outage capacity,
\begin{equation}
C^o_q = (1-q)C_q = (1-q) \log \left[ 1- P\bar{\gamma} \log(1-q)
\right], \label{eqn:outageCopt}
\end{equation}
defined as the rate averaged over outage and non-outage states
\cite{LiangJournal}, and provide $(q_C^*, R_{q_C^*})$ as the
interface to the source, where $q_C^*$ is the argument that
maximizes \eqref{eqn:outageCopt}. In fact the solution
\[
q^*_C = 1 - \exp \left\{ -
\frac{e^{W(P\bar{\gamma})}-1}{P\bar{\gamma}} \right\},
\]
with $W(z)$ the {\em Lambert-W} function solving $z = W(z)e^{W(z)}$,
is in general different from $q_D^*$ in \eqref{eqn:qDstar}.
It is insufficient for the channel to provide only $(q_C^*,
R_{q_C^*})$ as the interface; instead it should provide the entire
$(q, C_q)$ curve and let the source choose the optimal operating
point on this curve to minimize overall expected distortion. \\

{\em Separation schemes with broadcast channel code}: We have seen
in Section \ref{sec:channelMetrics} that a composite channel can be
viewed as a broadcast channel with virtual receivers indexed by each
channel state. A broadcast channel code can be applied to achieve
rate $R_s$ when channel is in state $s$. Since a Gaussian source is
successively refinable \cite{Equitz91} we can design a
multi-resolution source code which, when combined with the broadcast
channel code, achieves distortion $D(R_s)$ for each channel state
$s$. The overall expected distortion is $\mathbb{E}_SD(R_S)$.

We assume a power allocation profile $\rho(\gamma)\ge 0$ which
satisfies the overall power constraint $\int_0^\infty \rho(\gamma) d
\gamma = P$. It is shown in \cite{Shamai03} that the following rate,
in unit of nats per channel use, is achievable when the channel gain
is $\gamma$
\[
R(\gamma) = \int_0^\gamma \frac{u \rho(u)}{1+uI(u)} du.
\]
Here $I(\gamma) = \int_\gamma^\infty \rho(u) du $ is the
interference level for channel state $\gamma$. The minimum expected
distortion with a multi-resolution source code and a broadcast
channel code is then
\begin{equation}
\min_{\rho(\gamma)}  \int_0^\infty \sigma^2 e^{-R(\gamma)} p(\gamma)
d\gamma. \label{eqn:minD}
\end{equation}
The optimization problem \eqref{eqn:minD} was solved in \cite{Ng07b}
\cite{Tian07}. The optimal power allocation satisfies
\[
\rho_D^*(\gamma) = \left\{
\begin{array}{ll}
0, &  \gamma < \gamma_P \,\, \text{or} \,\, \gamma > \bar{\gamma}, \\
- I'(\gamma), &  \gamma_P \le \gamma \le \bar{\gamma}, \\
\end{array}
\right.
\]
where
\[
I(\gamma) = \frac{\int_{\bar{\gamma}}^\gamma
\left(\frac{1}{2\bar{\gamma}} - \frac{1}{u} \right)
e^{-u/2\bar{\gamma}} du } {\gamma e^{-\gamma/2\bar{\gamma}}},
\]
and $\gamma_P$ solves $I(\gamma_P) = P$. The minimum expected
distortion is
\[
D^e_2 = \sigma^2 \left[ D(\gamma_P) + \int_0^{\gamma_P} p(\gamma)
d\gamma \right],
\]
where
\[
D(\gamma) = \frac{ e^{-1} - \frac{1}{\bar{\gamma}}
\int_{\bar{\gamma}}^\gamma e^{-(u+\bar{\gamma})/2\bar{\gamma}}
\left(u/\bar{\gamma}\right)^{-1} du}
{\left(\gamma/\bar{\gamma}\right)^{-1}
e^{(\gamma-\bar{\gamma})/2\bar{\gamma}} }.
\]
In general the optimal power allocation $\rho_C^*(\gamma)$ that
maximizes the expected capacity $ \int_0^\infty R(\gamma) p(\gamma)
d\gamma$, as determined in \cite{Shamai03},
is different from $\rho_D^*(\gamma)$ that minimizes the expected
distortion \eqref{eqn:minD}. Therefore the channel should provide
the entire capacity region boundary $\{(R_s)_{s \in \mathcal{S}}\}$
as the interface.

In \refF{compare} we plot the expected distortion under the
different source-channel coding schemes, assuming average channel
gain $\bar{\gamma}=1$ and source variance $\sigma^2=1$. It is
observed that the broadcast channel code combined with the
multi-resolution source code performs slightly better than the
channel code for capacity versus outage combined with a single rate
source code, but there is a large gap between their expected
distortion and that of the optimal uncoded scheme.
\Fig{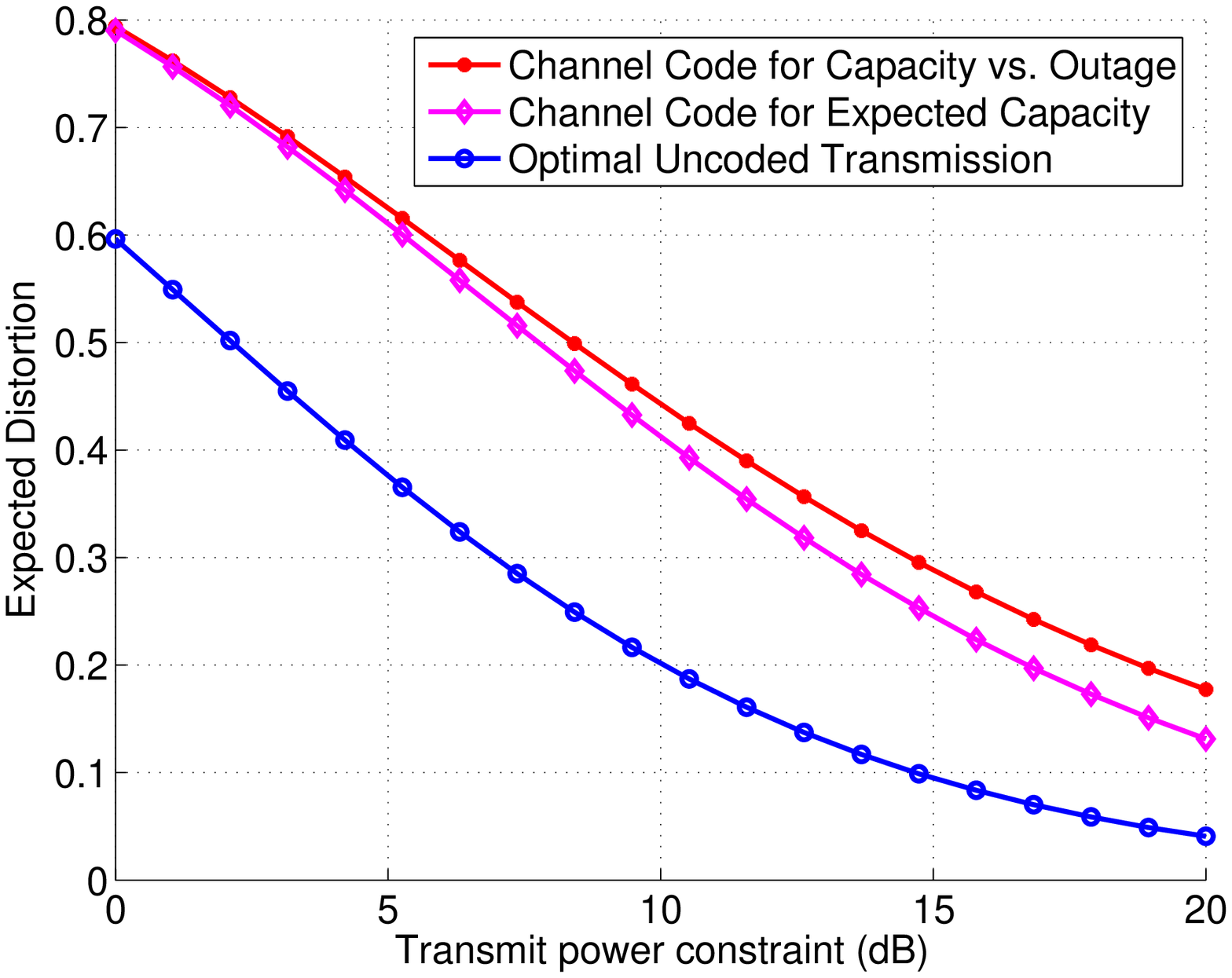}{Expected distortion for various source-channel coding
schemes}

\section{Source-Channel Interface under Expected Distortion Metric}
\label{sec:InterfaceBSC}

When the end-to-end performance metric is expected distortion,
separation schemes are usually suboptimal. In Section
\ref{sec:ExampleGaussianSourceSlowFadingChannel} we showed an
example of transmission of a Gaussian source over a slow fading
Gaussian channel. The uncoded transmission scheme is optimal if the
bandwidth expansion ratio $b=1$. With bandwidth compression or
expansion $(b\neq 1)$, various joint source-channel coding schemes
based on layering and hybrid analog-digital transmission
\cite{Shamai98, Reznic06, Mittal02} have been proposed to achieve
lower expected distortion than separation schemes. However, even the
simplest problem of transmitting a Gaussian source over a two-state
composite Gaussian channel is still open - so far no generally
optimal scheme is known.

For joint coding schemes, the concept of source-channel information
exchange through the interface still applies. Before transmission
starts, in separation schemes the source and channel exploit the
negotiation interface to agree on a single or a set of encoding
rates. In joint coding schemes, besides encoding rates, information
about other source and channel statistics may be exchanged. For
example, in hybrid digital-analog coding schemes \cite{Mittal02} the
channel provides the encoding rates for the digital part and the
channel bandwidth for the analog part as the negotiation interface.

After transmission starts, although we may not separate the
encoder/decoder into a source encoder/decoder and a channel
encoder/decoder for joint coding schemes, we can still identify a
source processing unit and a channel processing unit in many cases.
At the transmitter side, in contrast to that of a source encoder,
the output of a source processing unit is not necessarily from an
index set. For example, in a vector-quantization based joint coding
scheme \cite{Skoglund02}, the source processing unit provides both
the quantization index and residue to the channel processing unit
through the transmitter interface. Similarly at the receiver side,
the channel processing unit provides an estimate of the quantization
index and a noise-corrupted version of the quantization residue to
the destination processing unit through the receiver interface.

This notion of a source/channel processing unit is motivated by real
applications where the data collection and data transmission occur
in geographically dispersed locations. Sensor networks are one such
example, where sensor nodes obtain some local observations and
conduct some preliminary processing, and the processed data are then
delivered to remote fusion centers for long-haul transmission.  To
some extent this notion of source/channel processing unit is a
natural extension of source/channel encoder/decoder since it also
follows the philosophy of design by module; however, the flexibility
of separation is not retained - many schemes are tailored to the
specific system and are not universally applicable if the source or
channel is changed to other models.

Various source-channel coding schemes, separate or joint, can be
compared by their end-to-end expected distortions. The benefit of
many joint coding schemes comes at a price of more information
exchange through the interface. We believe a complete picture should
represent each scheme by a point on a two-dimensional plot, which
shows both end-to-end performance and interface complexity. The
choice of the transmission scheme then depends on the system
designer's view of the tradeoff between the two criterions. We
illustrate this methodology through the following example.

Consider transmission of a binary symmetric source over a two-state
composite BSC. Denote by $\alpha_i$, $i=1,2$, the random crossover
probability for each channel state. The two channel states occur
with probability $(1-p)$ and $p$, respectively. We assume $n$ source
bits are transmitted over $m$ channel uses and $m>n$, i.e. the
channel bandwidth expansion ratio $b = m/n > 1$. We also assume $0 <
\alpha_1 < \alpha_2 < (1/2)$ and $b [1 - h(\alpha_1)] < 1$, so even
the ``good" channel state $1$ cannot achieve lossless transmission.
The distortion measure between a source sequence and its
reconstruction is the Hamming distance
\[
d(V^n, \hat{V}^n) = \frac{1}{n} \sum_{\i=1}^n V_i \oplus \hat{V}_i.
\]

\subsection{Separate Source-Channel Coding}
\label{sec:SeparateMRBCcode} \psfrag{Vn}{$V^n$}
\psfrag{V1n}{$V^n_1$} \psfrag{V2n}{$V^n_2$} \psfrag{Q1n}{$Q^n_1$}
\psfrag{Q2n}{$Q^n_2$} \psfrag{Qbetam}{$Q^m_\beta$}
\psfrag{Um}{$U^m$} \psfrag{Xm}{$X^m$} \psfrag{Zm}{$Z^m$}
\psfrag{SOURCE}{Source} \psfrag{CHANNEL}{Channel} \psfrag{ENC 1}{ENC
1} \psfrag{ENC 2}{ENC 2} \psfrag{DEC 1}{DEC 1} \psfrag{DEC 2}{DEC 2}
\psfrag{Uhatm}{$\hat{U}^m$} \psfrag{Qbetahatm}{$\hat{Q}_\beta^m$}
\psfrag{Vhatn}{$\hat{V}^n$}
\psfrag{V2hatn}{$\hat{V}^n_2$}\psfrag{V1hatn}{$\hat{V}^n_1$}
\psfrag{SOURCE-CHANNEL INTERFACE}{Source-Channel Interface}
\psfrag{(TRANSMITTER)}{(transmitter)}
\psfrag{(RECEIVER)}{(receiver)}
\begin{figure}[htbp]
\begin{center}
\includegraphics
[width=4in]{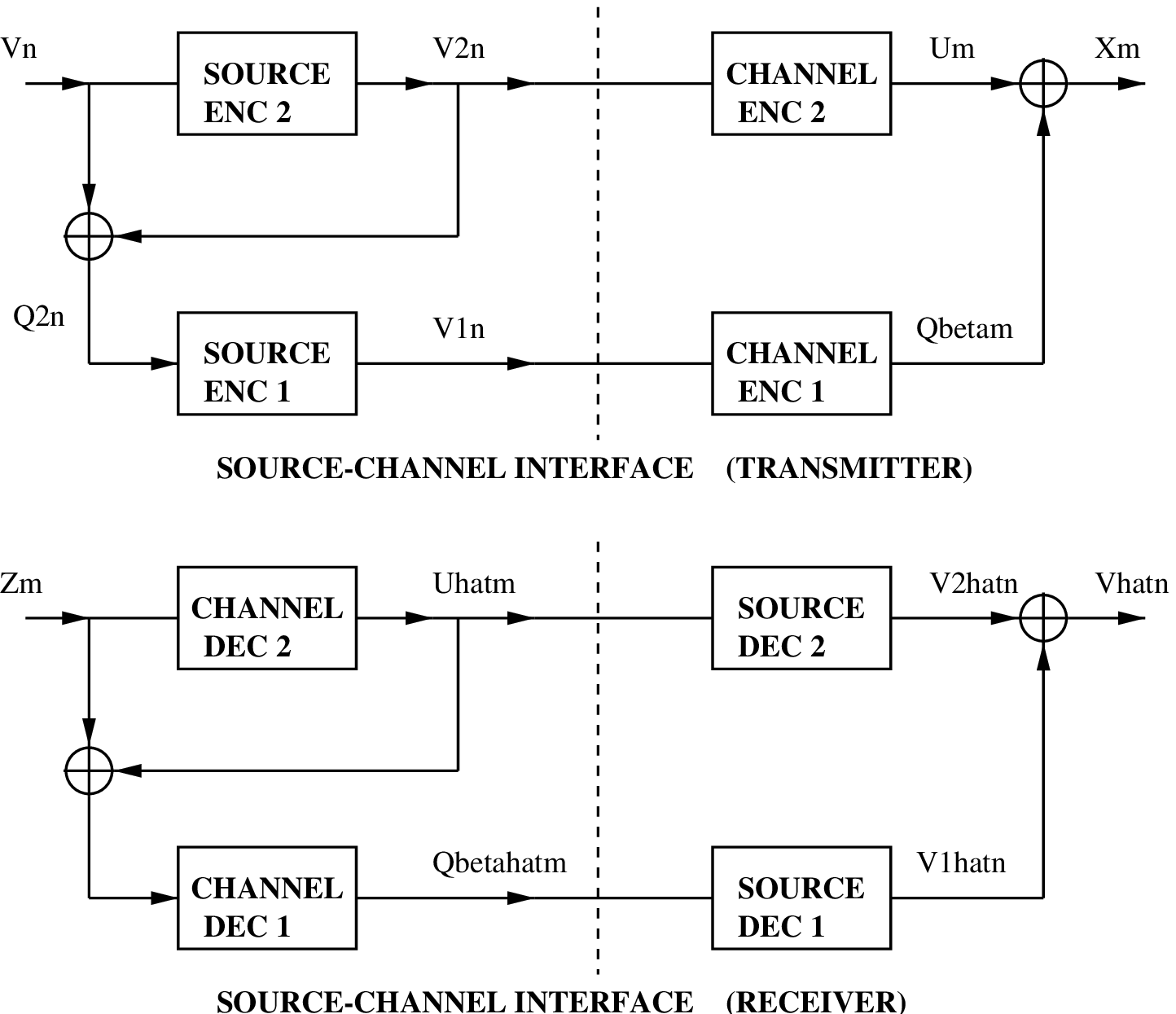}
\caption{Separate coding scheme. MR source code with BC channel
code.} \label{fig:expected_enc}
\end{center}
\end{figure}
The two states of the composite BSC have a degraded relationship and
can be viewed as two virtual receivers of a BSC-BC. The following
rate pairs, in unit of bits per channel use, are achievable using a
broadcast channel code \cite[p.425]{CoverIT}
\begin{eqnarray}
R_1 &\le& h(\alpha_1 * \beta) - h(\alpha_1), \nonumber \\
R_2 &\le& 1 - h(\alpha_2 * \beta), \label{eqn:MRBCRateCondition}
\end{eqnarray}
where $\alpha * \beta = \alpha (1-\beta) + \beta (1- \alpha)$, and
$h(\alpha) = - \alpha \log \alpha - (1-\alpha) \log (1- \alpha)$ is
the binary entropy function. The subscript $(\cdot)_2$ denotes the
common information that can be decoded in both states, and the
subscript $(\cdot)_1$ denotes the individual information that is
decodable only in the good state. By varying $\beta$ between $0$ and
$1/2$ we can trace the entire BC capacity region boundary.

Since a binary symmetric source is successively refinable under the
Hamming distortion measure \cite{Equitz91}, we can match the BC code
with a multi-resolution source code to achieve distortions
\begin{eqnarray}
D_1 &=& D\left( b(R_1 + R_2) \right), \nonumber \\
D_2 &=& D(bR_2) \label{eqn:MRBCDistortionCondition}
\end{eqnarray}
for each state, where $b$ is the bandwidth expansion ratio and
$D(R)$ is the distortion-rate function of a BSS, i.e. the inverse
function of $R(D) = 1 - h(D)$. The overall expected distortion is
given by
\[
D^e_{\BC} = (1-p) D_1 + p D_2.
\]
In \refF{expected_enc} we show the block diagram of this separate
source-channel coding scheme. The broadcast channel code has a
structure of additive superposition encoding and successive decoding
with interference cancellation \cite[p.379]{CoverIT}. The
multi-resolution source code is implemented as a multistage vector
quantization (MSVQ) \cite{Gersho92}. Using the {\em test channel}
interpretation of rate-distortion theory \cite[p.343]{CoverIT}, we
see that in the first stage, Source ENC 2 quantizes the source
sequence $V^n$ by $V^n_2$ and the residue $Q_2^n = V^n \oplus V_2^n$
is a Bernoulli$(D_2)$ sequence. In the second stage, Source ENC 1
further quantizes $Q_2^n$ by $V_1^n$ and the residue $Q_1^n = Q_2^n
\oplus V_1^n$ follows a Bernoulli$(D_1)$ distribution. Details about
the structure of the MR source code and BC code are given in
Appendix \ref{app:SCCodeForSeparationSchemes}.
\begin{table}[htbp]
\caption{Interface for separation scheme: Multi-resolution source
code and broadcast channel code} \centering
\begin{tabular}{l|p{2.6in}}
\hline Negotiation & achievable distortion with MR source code
$(D_1, D_2)$, BC capacity region $(R_1,R_2)$, channel state
probability $p$ \\
\hline Transmitter & $\mathcal{M}_{m,t} =
\{1,\cdots, 2^{mR_1}\} \times \{1,\cdots, 2^{mR_2}\} $ \\
\hline Receiver & $\mathcal{M}_{m,1} = \mathcal{M}_{m,t}$ for
channel state 1,  $\mathcal{M}_{m,2} =
\{1,2,\cdots, 2^{mR_2}\}$ for channel state 2.\\
\hline
\end{tabular}
\label{table:InterMRBC}
\end{table}

The interface of this scheme is summarized in Table
\ref{table:InterMRBC}, i.e. Table
\ref{table:ExpectedDistortionInterface} specified to the current
example. In \refF{expected_enc} the dashed lines clearly separate
the source and channel coders and identify the transmitter and
receiver interface. To measure the interface complexity, we consider
the number of bits per source symbol that are delivered through the
interface. The complexity of the transmitter interface is
\[
K^t_{\BC} = b(R_1 + R_2),
\]
and the receiver interface complexity is the expected capacity
multiplied by the bandwidth expansion ratio
\[
K^r_{\BC} = b[(1-p) R_1 + R_2].
\]

The separation scheme based on Shannon capacity is a special case
when $\beta = 0$. As a result, $R_2 = 1-h(\alpha_2)$ and $R_1 = 0$.
We only transmit the base layer information and achieve distortion
$D_1 = D_2 = D(bR_2)$ in both states. The transmitter and receiver
interface complexity is $K^t_{\textrm{Shannon}} =
K^r_{\textrm{Shannon}} = bR_2$ bits per channel use.

Similarly, when $\beta = 1/2$ we have the separation scheme based on
capacity versus outage. Here $R_1 = 1 - h(\alpha_1)$ and $R_2 = 0$.
We only transmit the refinement layer and achieve distortion $D_1 =
D(bR_1)$, $D_2 = (1/2)$. The transmitter interface complexity is
$K^t_{\textrm{outage}} = bR_1$, and the receiver interface
complexity is $K^r_{\textrm{outage}} = (1-p)bR_1$, which is
proportional to the outage capacity.

\subsection{Systematic Coding}
\psfrag{WynerZiv}{Wyner-Ziv} \psfrag{ENC}{ENC} \psfrag{DEC}{DEC}
\psfrag{iVWZ}{$j(\tilde{V}^n)$} \psfrag{Xm-n}{$X^{m-n}$}
\psfrag{Zm-n}{$Z^{m-n}$} \psfrag{Xm-n+1:m}{$X^{m-n+1:m}$}
\psfrag{PRIMARY}{Primary} \psfrag{SECONDARY}{Secondary}
\psfrag{Zm-n+1:m}{$Z^{m-n+1:m}$}
\psfrag{hatiVWZ}{$\hat{j}(\tilde{V}^n)$}
\psfrag{hatVWZ}{$\hat{\tilde{V}}^n$} \psfrag{hatVu=}{$V_u^n=$}
\begin{figure}[htbp]
\begin{center}
\includegraphics
[width=4in]{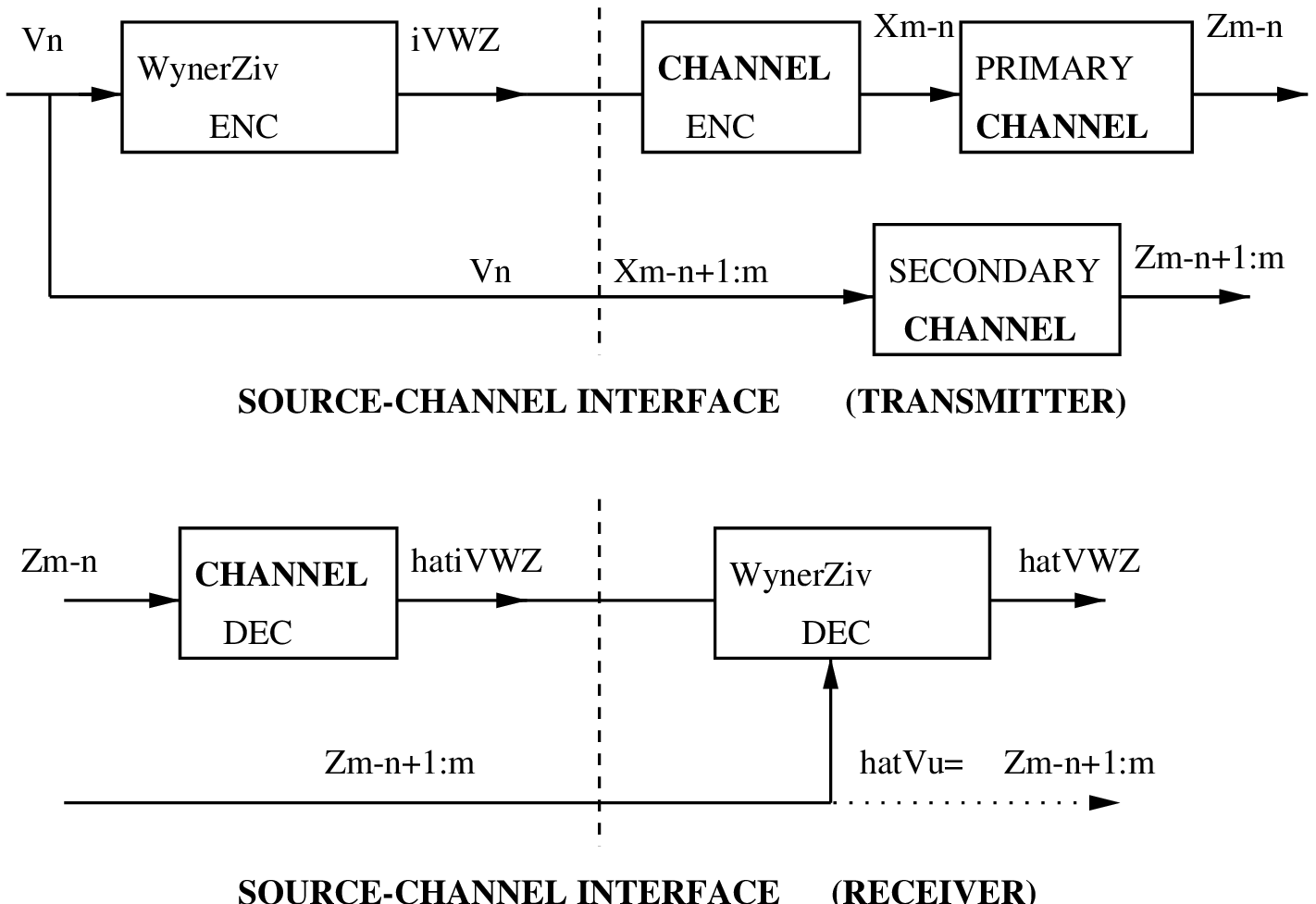} \caption{Systematic coding scheme.}
\label{fig:wz_enc}
\end{center}
\end{figure}

Recall that $n$ source bits are transmitted in $m$ channel uses and
we assume $m>n$. The channel is divided into a primary channel and a
secondary channel. The uncoded $n$ source bits are directly
transmitted over the secondary channel in $n$ channel uses. The
output of the secondary channel provides side information about the
source sequence at the destination. We then apply the Wyner-Ziv code
\cite{WynerZiv76}, which is a source coding technique with side
information at the decoder, and transmit the encoder output over the
primary channel in the remaining $(m-n)$ channel uses. The name {\em
systematic coding} comes from its similarity to the {\em systematic
linear block code} \cite[p.85]{Wicker}, where
the input information bits are embedded in the output codewords.
This scheme is motivated by \cite{Shamai98}.

The rate-distortion function for Wyner-Ziv coding with side
information is given by \cite{WynerZiv76}
\begin{equation}
R^*(d) = \left\{
\begin{array}{ll}
g(d), & 0 \le d \le d_c, \\
g(d_c) {\displaystyle \frac{\alpha-d}{\alpha - d_c} } = - g'(d_c)
(\alpha-d), & d_c < d \le \alpha,
\end{array}
\right. \label{eqn:WynerZivRateDistortion}
\end{equation}
where $\alpha$ is the BSC crossover probability, the function $g(d)$
is defined as
\[
g(d) = \left\{
\begin{array}{ll}
h(\alpha*d) - h(d), & 0 \le d < \alpha, \\
0, & d = \alpha,
\end{array}
\right.
\]
$g'(d)$ is the derivative of $g(d)$, and the turning point $d_c$ is
the solution to
\begin{equation}
\frac{g(d_c)}{d_c - \alpha} = g'(d_c). \label{eqn:dcDefn}
\end{equation}
We give a brief review of the achievability of the rate-distortion
function $R^*(d)$. Notice that $R^*(\alpha) = 0$ is achievable by
simply observing the side information, i.e. the secondary channel
output due to the uncoded source bits. We focus on the case of $0
\le d \le d_c$. For $ d_c < d \le \alpha $, $R^*(d)$ is achievable
by time sharing between $(\alpha, 0)$ and $(d_c, R^*(d_c))$.
Basically, for a source sequence $V^n$ drawn i.i.d. from a
Bernoulli$(1/2)$ distribution, the output of the secondary channel
is
\[
V_u^n = V^n \oplus Q_\alpha^n,
\]
where the channel noise $Q_\alpha^n$ is an i.i.d.
Bernoulli$(\alpha)$ sequence. The Wyner-Ziv codebook $\mathcal{C}$
consists of $2^{n[1-h(d)]}$ codewords $\tilde{V}^n$, drawn i.i.d.
from a Bernoulli$(1/2)$ distribution. We can approximate each source
sequence $V^n$ by a quantized version $\tilde{V}^n$ with residue
$Q_d^n$, i.e.
\[
V^n = \tilde{V}^n  \oplus Q_d^n.
\]
Using the {\em test channel} concept of rate-distortion theory
\cite[p.343]{CoverIT}, $Q_d^n$ is an i.i.d. Bernoulli$(d)$ sequence
independent of $\tilde{V}^n$. We want to recover $\tilde{V}^n$ at
the destination in order to estimate the source sequence $V^n$
within distortion $d$. Without side information, we have to transmit
the index of each $\tilde{V}^n$ using $\log |\mathcal{C}| =
n[1-h(d)]$ bits. On the other hand, the secondary channel output
\[
V_u^n = V^n \oplus Q_\alpha^n = \tilde{V}^n  \oplus Q_d^n \oplus
Q_\alpha^n
\]
also provides information about $\tilde{V}^n$ in terms of $I(V^n_u;
\tilde{V}^n) = n[1-h(\alpha * d)]$. Using the random binning
technique \cite[p.411]{CoverIT}, we can uniformly distribute the
$\tilde{V}^n$ sequences into
\[
\frac{2^{n[1-h(d)]}}{2^{n[1-h(\alpha * d)]}} = 2^{n[h(\alpha
* d)-h(d)]}
\]
bins, transmit the bin index $\hat{j}(\tilde{V}^n)$ instead of the
sequence index, and hence reduce the encoding rate from $1-h(d)$ to
$h(\alpha * d)-h(d)$. With receiver side information the sequence
$\tilde{V}^n$ can still be decoded with small error. This approach
is formalized in \cite[Sec. II]{WynerZiv76}.

The Wyner-Ziv coding rate depends on the quality of the side
information, i.e. the BSC crossover probability $\alpha$. We can
construct two systematic codes, one for each channel state $\alpha =
\alpha_i$, $i = 1,2$. For the systematic code targeting the good
channel state, if the channel is indeed in the good state, we can
decode the Wyner-Ziv code with side information $V_u^n$ and the
achievable distortion is determined by
\[
R_1^*(D_1) = (b-1) C_1 = (b-1) [1 - h(\alpha_1)],
\]
where $C_1$ is the channel capacity for good state, $R_1^*(d)$ is
the rate-distortion function \eqref{eqn:WynerZivRateDistortion} with
$\alpha = \alpha_1$. Note
that 
this information is transmitted over the primary channel with
bandwidth expansion ratio $(b-1)$, since it only consists of $m-n =
(b-1)n$ channel uses. If the channel is actually in the bad state,
we cannot decode the Wyner-Ziv code. Instead we estimate the source
by the secondary channel output and achieve a distortion $D_2 =
\alpha_2$.
\begin{table}[htbp]
\caption{Interface for systematic coding scheme targeting the good
channel state} \centering
\begin{tabular}{l|p{2.6in}}
\hline Negotiation & Wyner-Ziv rate-distortion function $R_1^*(d)$,
primary channel capacity $C_1$, secondary channel statistics ($n$
uses of BSC), channel state probability $p$ \\
\hline Transmitter & uncoded source sequence $V^n$, Wyner-Ziv
encoder output $\mathcal{M}_{m-n,t} =
\{1,2,\cdots, 2^{(m-n)C_1}\}$ \\
\hline Receiver & secondary channel output $V_u^n$ for both states,
$\mathcal{M}_{m-n,1} = \mathcal{M}_{m-n,t}$ for channel state 1
only. \\
\hline
\end{tabular}
\label{table:InterfaceSystematicGood}
\end{table}

The interfaces of this scheme are summarized in Table
\ref{table:InterfaceSystematicGood}, and are also illustrated by the
dashed lines in \refF{wz_enc}. The interfaces divide the source and
the channel processing units so that we can still design by module,
but these processing units are no longer categorized as source or
channel coders because of the uncoded transmission over the
secondary channel. Similar to previous separation schemes, we
measure the interface complexity by the number of bits per source
symbol that are delivered through the interface. The complexity of
the transmitter interface is
\[
K^t_{\textrm{SYS,1}} = 1 + (b-1)[1-h(\alpha_1)],
\]
and the complexity of the receiver interface is
\[
K^r_{\textrm{SYS,1}} = 1 + (1-p)(b-1)[1-h(\alpha_1)].
\]

Similarly we can construct a systematic code targeting the bad
channel state. If the channel is indeed in state 2, the achievable
distortion $D_2$ is determined by
\begin{equation}
R_2^*(D_2) = (b-1) [1 - h(\alpha_2)], \label{eqn:R2*D2}
\end{equation}
where $R_2^*(d)$ is the rate-distortion function
\eqref{eqn:WynerZivRateDistortion} with $\alpha = \alpha_2$. If the
channel is in the good state, we have different options:
\begin{itemize}
\item $D_2 \le d_{c2}$, where 
$d_{c2}$ is the turning point given by \eqref{eqn:dcDefn}. Here the
source code does not involve any time-sharing. The quality of the
side information is actually better than targeted so we can also
perform Wyner-Ziv decoding, recover $\tilde{V}^n$, and reconstruct
the source within distortion $D_2$. Or we can simply observe the
secondary channel output and achieve a distortion of $\alpha_1$.
Therefore $D_1 = \min\{D_2, \alpha_1\}$.

\item $D_2 > d_{c2}$. Here the source code involves a time sharing between
the uncoded transmission and the Wyner-Ziv code with distortion
$d_{c2}$. The time sharing factor $\theta$ is determined by
\[
D_2 = \theta d_{c2} + (1-\theta) \alpha_2.
\]
In the good state, for proportion $(1-\theta)$ of the time, we use
the secondary channel output and achieve a distortion of $\alpha_1$.
For proportion $\theta$ of the time, we can reconstruct the source
from the Wyner-Ziv code or the secondary channel output, and achieve
a distortion of $\min\{d_{c2}, \alpha_1\}$. The overall distortion
after time-sharing becomes
\[
D_1 = \theta \min\{d_{c2}, \alpha_1\} + (1-\theta) \alpha_1.
\]
\end{itemize}
The above two cases can be combined as follows
\[
D_1 = \left\{
\begin{array}{ll}
\alpha_1, & \alpha_1 \le \min\{D_2, d_{c2}\}, \\
D_2, & D_2 \le d_{c2}, D_2 < \alpha_1, \\
\theta d_{c2} + (1-\theta) \alpha_1, & d_{c2} < D_2, d_{c2} <
\alpha_1.
\end{array}
\right.
\]
The complexity of the transmitter interface is
\[
K^t_{\textrm{SYS,2}} = 1 + (b-1)[1-h(\alpha_2)],
\]
and the complexity of the receiver interface is
\[
K^r_{\textrm{SYS,2}} = \left\{
\begin{array}{ll}
1 + p(b-1)[1-h(\alpha_2)], & \alpha_1 \le \min\{D_2, d_{c2}\}, \\
1 + (b-1)[1-h(\alpha_2)], & \alpha_1 > \min\{D_2, d_{c2}\}, \\
\end{array}
\right.
\]
i.e. for the good channel state we perform Wyner-Ziv decoding if and
only if $\alpha_1 > \min\{D_2, d_{c2}\}$

\subsection{Quantization Residue Splitting}
\psfrag{Demux}{Demux} \psfrag{V11-rn}{$V_1^{(1-\rho)n}$}
\psfrag{Q21-rn}{$Q_2^{(1-\rho)n}$}
\psfrag{Q21-rn+1:n}{$Q_2^{(1-\rho)n+1:n}$} \psfrag{Um-rn}{$U^{m-\rho
n}$} \psfrag{Xm-rn}{$X^{m-\rho n}$}\psfrag{Qbm-rn}{$Q_\beta^{m-\rho
n}$} \psfrag{Zm-rn}{$Z^{m-\rho n}$} \psfrag{Xm-rn+1:m}{$X^{m-\rho
n+1:m}$}\psfrag{Zm-rn+1:m}{$Z^{m-\rho n+1:m}$}
\psfrag{hatUm-rn}{$\hat{U}^{m-\rho
n}$}\psfrag{hatQbm-rn}{$\hat{Q}_\beta^{m-\rho n}$}
\psfrag{hatQ21-rn}{$\hat{V}_1^{(1-\rho)n}$}
\psfrag{hatV1-rn}{$\hat{V}^{(1-\rho)n}$}
\psfrag{hatV1-rn+1:n}{$\hat{V}^{(1-\rho)n+1:n}$}
\psfrag{SOURCE-CHANNEL INTERFACE (TRANS)}{Source-Channel Interface
(transmitter)} \psfrag{SOURCE-CHANNEL INTERFACE
(REC)}{Source-Channel Interface (receiver)}
\begin{figure*}[htbp]
\begin{center}
\includegraphics
[width=6.7in]{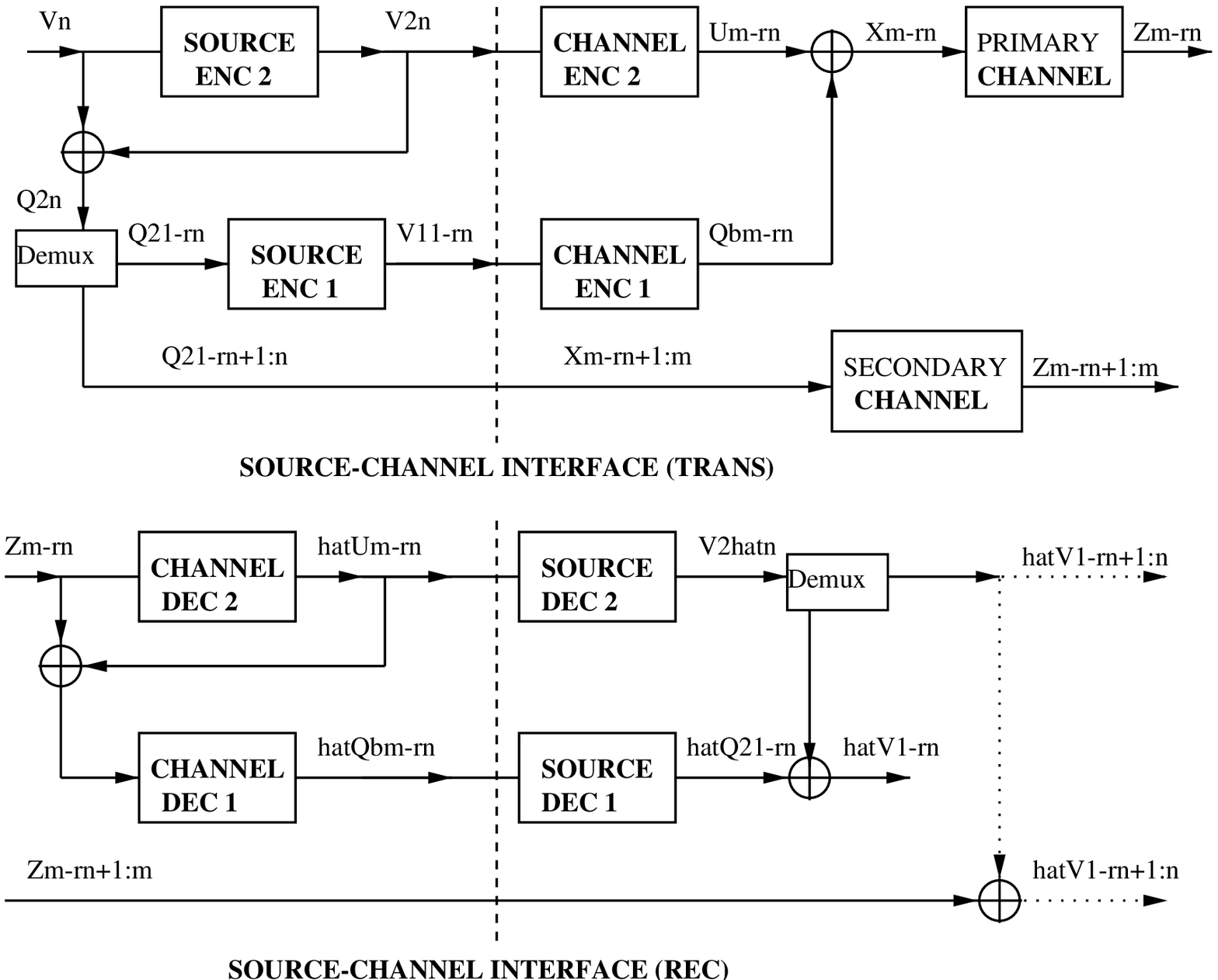} \caption{Quantization residue
splitting scheme.} \label{fig:Quantization_residue_splitting_enc}
\end{center}
\end{figure*}

The block diagram of this coding scheme is shown in
\refF{Quantization_residue_splitting_enc}. The overall channel is
divided into two subchannels, a secondary channel of $\rho n$
channel uses, $0 \le \rho \le 1$, and a primary channel of the
remaining $ m - \rho n = (b-\rho) n$ channel uses. For the primary
channel, we use the same BC code as in Section
\ref{sec:SeparateMRBCcode} to achieve the rate pair
\eqref{eqn:MRBCRateCondition}
\begin{eqnarray*}
R_1 &\le& h(\alpha_1 * \beta) - h(\alpha_1), \\
R_2 &\le& 1 - h(\alpha_2 * \beta).
\end{eqnarray*}
Similar to the MR code in Section \ref{sec:SeparateMRBCcode}, we
first quantize the source sequence $V^n$ at rate $(b-\rho)R_2$. Note
that the bandwidth expansion ratio for the primary channel is
$(b-\rho)$. The quantization output $V^n_2$ is to be decoded in both
channel states. The quantization residue $Q_2^n = V^n \oplus V_2^n$
follows a Bernoulli$(d_2)$ distribution with
\[
d_2 = D((b-\rho) R_2).
\]
We then split the residue into two sequences, $Q_2^{(1-\rho)n}$ of
the first $(1-\rho)n$ bits, and $Q_2^{(1-\rho)n+1:n}$ of the
remaining $\rho n$ bits. The sequence $Q_2^{(1-\rho)n}$ is quantized
at rate $\frac{b-\rho}{1-\rho}R_1$. The output $V_1^{(1-\rho)n}$ is
to be decoded by channel state $1$ only, and it is superimposed over
the first-stage quantization output $V^n_2$ and transmitted over the
primary channel using the previous BC code. The sequence
$Q_2^{(1-\rho)n+1:n}$ is directly transmitted over the secondary
channel, and the channel output is
\[
Z^{m-\rho n+1:m} = Q_2^{(1-\rho)n+1:n} \oplus
Q_{\alpha}^{(1-\rho)n+1:n},
\]
where $Q_{\alpha}^{(1-\rho)n+1:n}$, $\alpha = \alpha_i$, $i=1,2$ is
the channel noise for each state. The separation scheme in Section
\ref{sec:SeparateMRBCcode} can be viewed as the special case of
$\rho = 0$. Extension to the current residue splitting scheme is
motivated by \cite{Mittal02}.

In the good channel state, the first $(1 - \rho)n$ bits are
reconstructed by decoding both layers, i.e.
\[
\hat{V}^{(1-\rho)n} = \hat{V}_1^{(1-\rho)n} \oplus
\hat{V}_2^{(1-\rho)n}.
\]
The achievable distortion is
\[
d_1 = D\left( \frac{b-\rho}{1-\rho}R_1 + (b - \rho) R_2 \right).
\]
The remaining $\rho n$ bits can be reconstructed by either the first
layer only, i.e. $\hat{V}_2^{(1-\rho)n+1:n}$, to achieve a
distortion of $d_2$, or further combined with the secondary channel
output, i.e.
\begin{eqnarray*}
&& \hat{V}^{(1-\rho)n+1:n} \\
&=& \hat{V}_2^{(1-\rho)n+1:n} \oplus Z^{m-\rho n+1:m} \\
&=& \hat{V}_2^{(1-\rho)n+1:n} \oplus Q_2^{(1-\rho)n+1:n} \oplus
Q_{\alpha 1}^{(1-\rho)n+1:n}
\end{eqnarray*}
to achieve a distortion of $\alpha_1$. The overall achievable
distortion for the good state is
\[
D_1 = (1-\rho) d_1 + \rho \min \{ d_2, \alpha_1 \}.
\]
In the bad channel state, we cannot decode the refinement layer and
the reconstruction by the base layer only achieves a distortion of
$d_2$. However, for the last $\rho n$ bits, we can also combine the
base layer decoding output
with the secondary channel output to achieve a
distortion of $\alpha_2$. Therefore the overall achievable
distortion for the bad state is
\[
D_2 = (1-\rho) d_2 + \rho \min \{ d_2, \alpha_2 \}.
\]

The interfaces of this scheme is summarized in Table
\ref{table:InterfaceResidueSplitting} and illustrated by the dashed
lines in \refF{Quantization_residue_splitting_enc}. The complexity
of the transmitter interface, measured as the number of bits per
source symbol delivered through the interface, is equal to
\[
K^t_{\textrm{QRS}} = (b-\rho) (R_1 + R_2) + \rho,
\]
where the subscript $(\cdot)_{\textrm{QRS}}$ denotes quantization
residue splitting. The complexity of the receiver interface is
\[
K^r_{\textrm{QRS}} = \left\{
\begin{array}{ll}
(b-\rho) [(1-p)R_1 + R_2], & d_2 \le \alpha_1, \\
(b-\rho) [(1-p)R_1 + R_2] + (1-p) \rho, & \alpha_1 < d_2 \le \alpha_2, \\
(b-\rho) [(1-p)R_1 + R_2] + \rho,  & d_2 > \alpha_2, \\
\end{array}
\right.
\]
i.e., for the primary channel the base layer output is delivered in
both states and the refinement layer only in channel state $1$. The
secondary channel output is delivered to the destination processing
unit in state $i$, if $d_2 > \alpha_i$.
\begin{table}[htbp]
\caption{Interface for quantization residue splitting scheme}
\centering
\begin{tabular}{l|p{2.8in}}
\hline Negotiation & rate-distortion pair $(d_1, d_2)$ for the MR
source code, primary channel BC capacity region $(R_1, R_2)$,
secondary channel statistics ($\rho n$ uses of BSC),
channel state probability $p$ \\
\hline Transmitter & uncoded partial quantization residue sequence
$Q_2^{(1-\rho)n+1:n}$, $\mathcal{M}_{m- \rho n,t} = \{1,\cdots,
2^{(m-\rho n)R_1}\} \times
\{1,\cdots, 2^{(m-\rho n)R_2}\} $ \\
\hline Receiver & $\mathcal{M}_{m-\rho n,1} = \mathcal{M}_{m-\rho
n,t}$ for channel state 1, $\mathcal{M}_{m-\rho n,2} = \{1,\cdots,
2^{(m-\rho n)R_2}\}$ for channel state 2, secondary channel output
$Z^{m-\rho n+1:m}$ for channel state $i$ if $d_2 > \alpha_i$,
$i=1,2$. \\
\hline
\end{tabular}
\label{table:InterfaceResidueSplitting}
\end{table}

\subsection{Numerical Examples}
We provide some numerical examples to compare different schemes in
this section. We assume the two states of the composite BSC have
crossover probabilities $\alpha_1 = 0.25$ and $\alpha_2 = 0.45$, and
the bandwidth expansion ratio $b = 2$.

\Fig{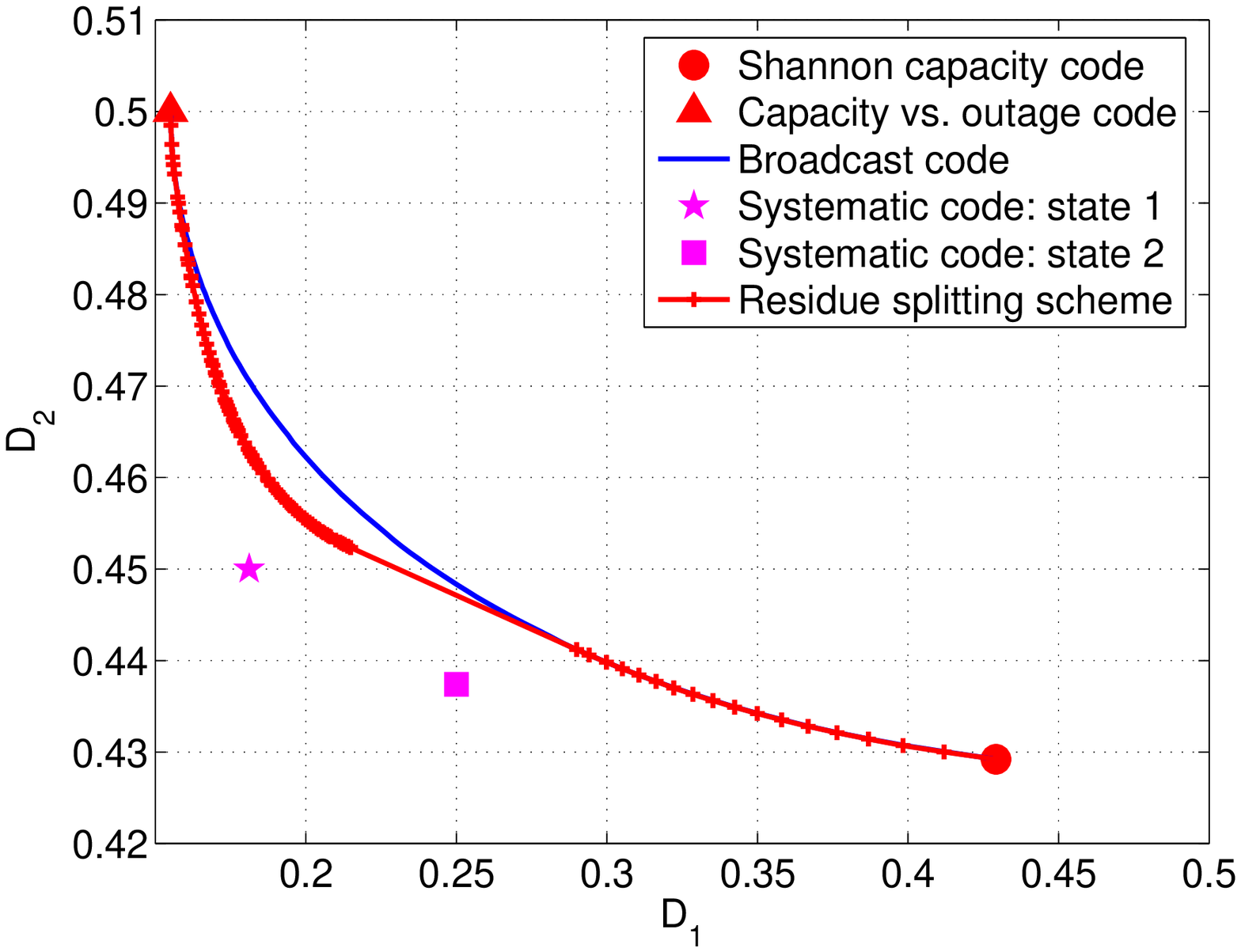}{Achievable distortion region $(D_1, D_2)$ for various
schemes.} In \refF{D1D2} we plot the achievable distortion pair
$(D_1, D_2)$ for each scheme. For the broadcast coding scheme, by
varying the auxiliary variable $\beta$ from $0$ and $1/2$, we change
the rate allocation between the base layer $(R_2)$ and the
refinement layer $(R_1)$. The separation schemes using the Shannon
capacity code and the capacity versus outage code are the special
cases of $\beta = 0$ and $1/2$, respectively. They are marked by the
two end-points of the broadcast distortion region boundary. For the
quantization residue splitting scheme, we calculate the distortion
pairs $(D_1, D_2)$ for different parameters $0 \le \beta \le 1/2$
and $0 \le \rho \le 1$. The plotted curve is the convex hull of all
achievable distortion pairs. Note that the broadcast scheme is a
special case of the residue splitting scheme with $\rho = 0$, so the
broadcast distortion region lies strictly within the residue
splitting distortion region. There are two systematic codes, one
targeting at each channel state. They are represented by two points,
both out of the residue splitting distortion region.

\Fig{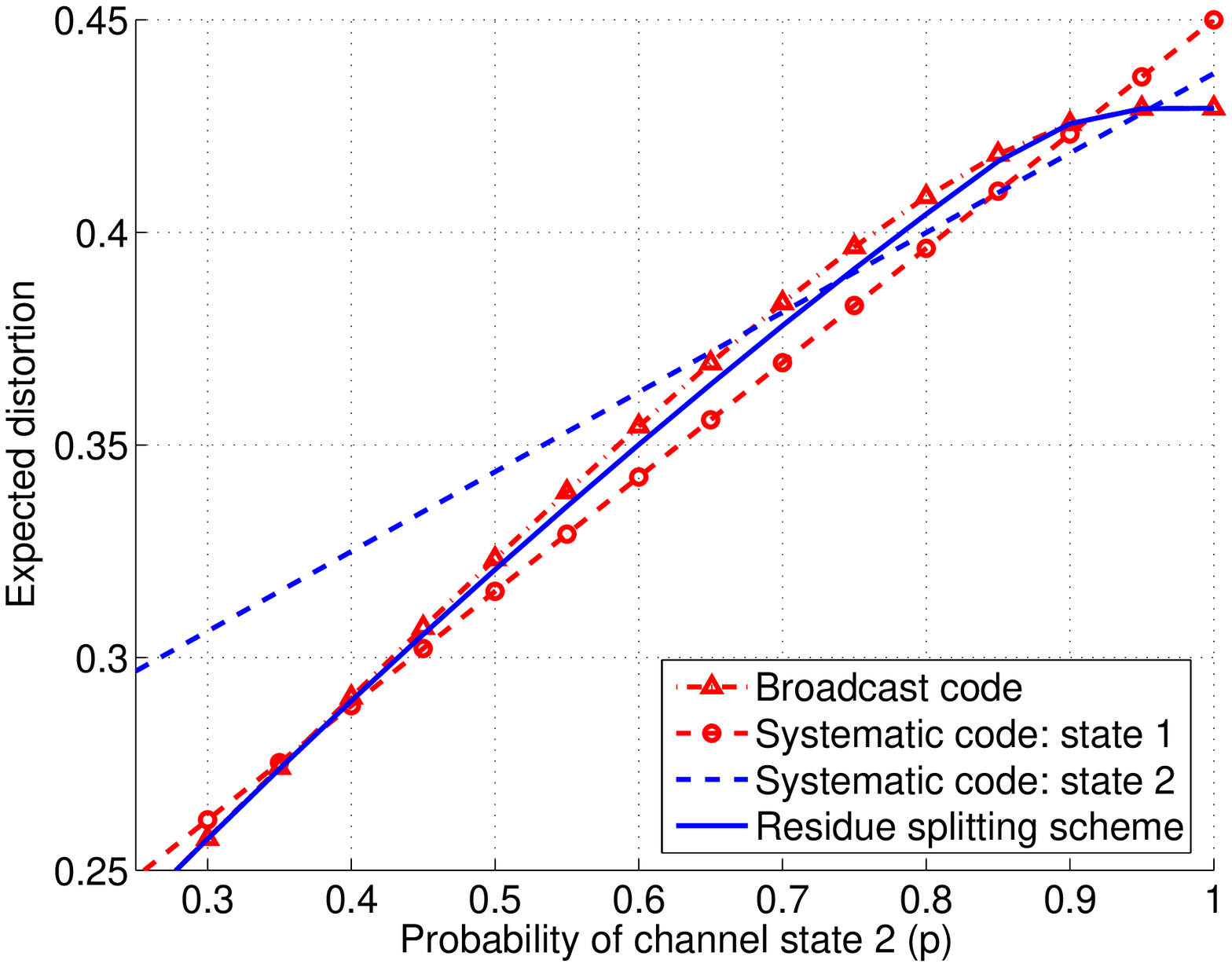}{Expected distortion for various channel state
distributions.} In \refF{EDcompare} we plot the expected distortion
of various schemes for different channel state distributions. Each
systematic code achieves a single distortion pair, so the expected
distortion is simply the weighted average and increases linearly
with the bad channel state probability $p$. For broadcast and
residue splitting schemes, we need to choose the optimal point on
the distortion region boundary at each channel state probability.
Since the broadcast scheme is a special case of the residue
splitting scheme, its expected distortion is no less, and sometimes
strictly larger, than that of the residue splitting scheme. For
different ranges of $p$, the scheme that achieves the lowest
expected distortion is also different. For $p<0.378$ or $p>0.956$ it
is the residue splitting scheme, for $0.378 < p < 0.845$ it is the
systematic code for the good channel state, and for $0.845 < p <
0.956$ it is the systematic code for the bad channel state.

\Fig{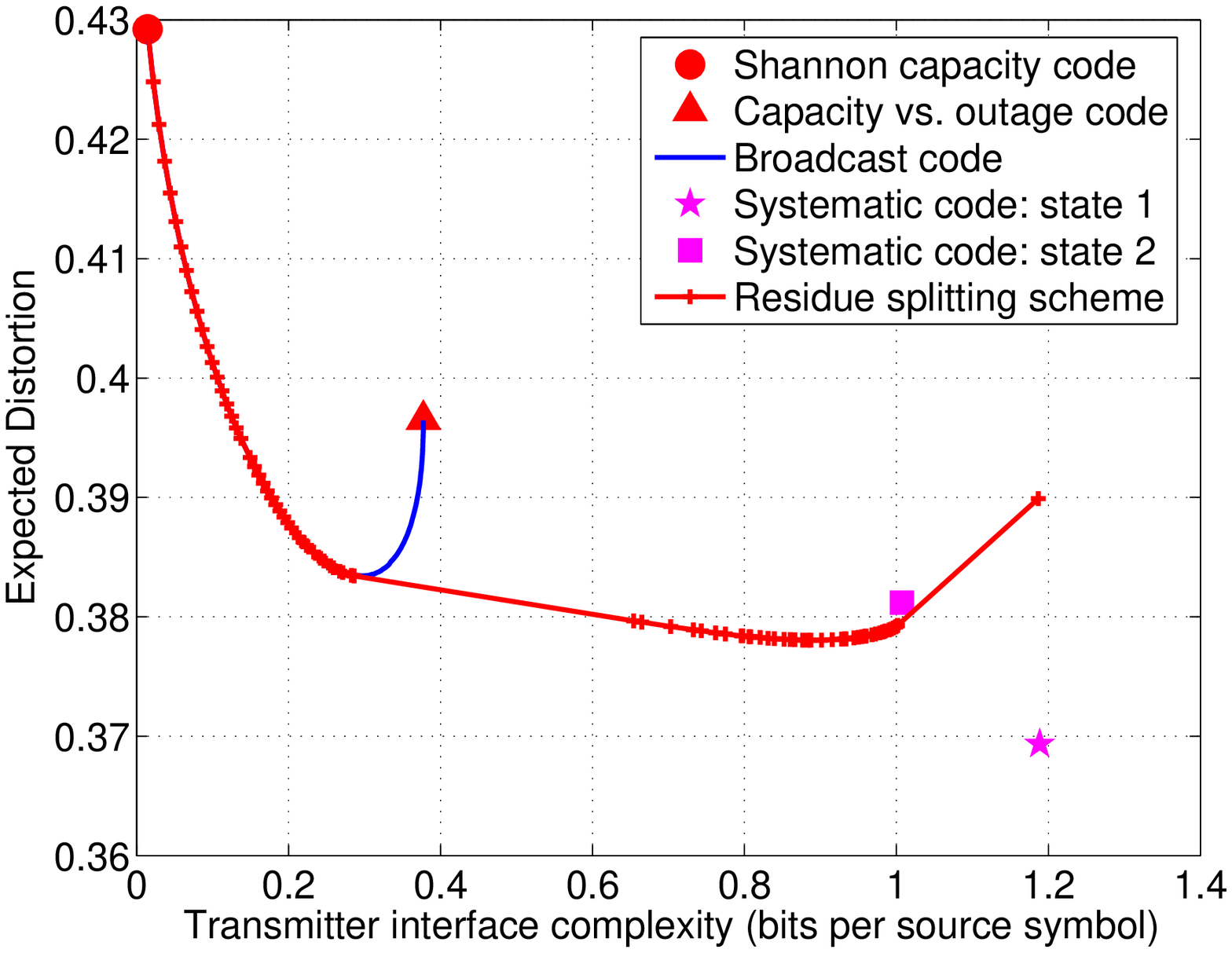}{Transmitter interface complexity vs. expected
distortion tradeoff.} \Fig{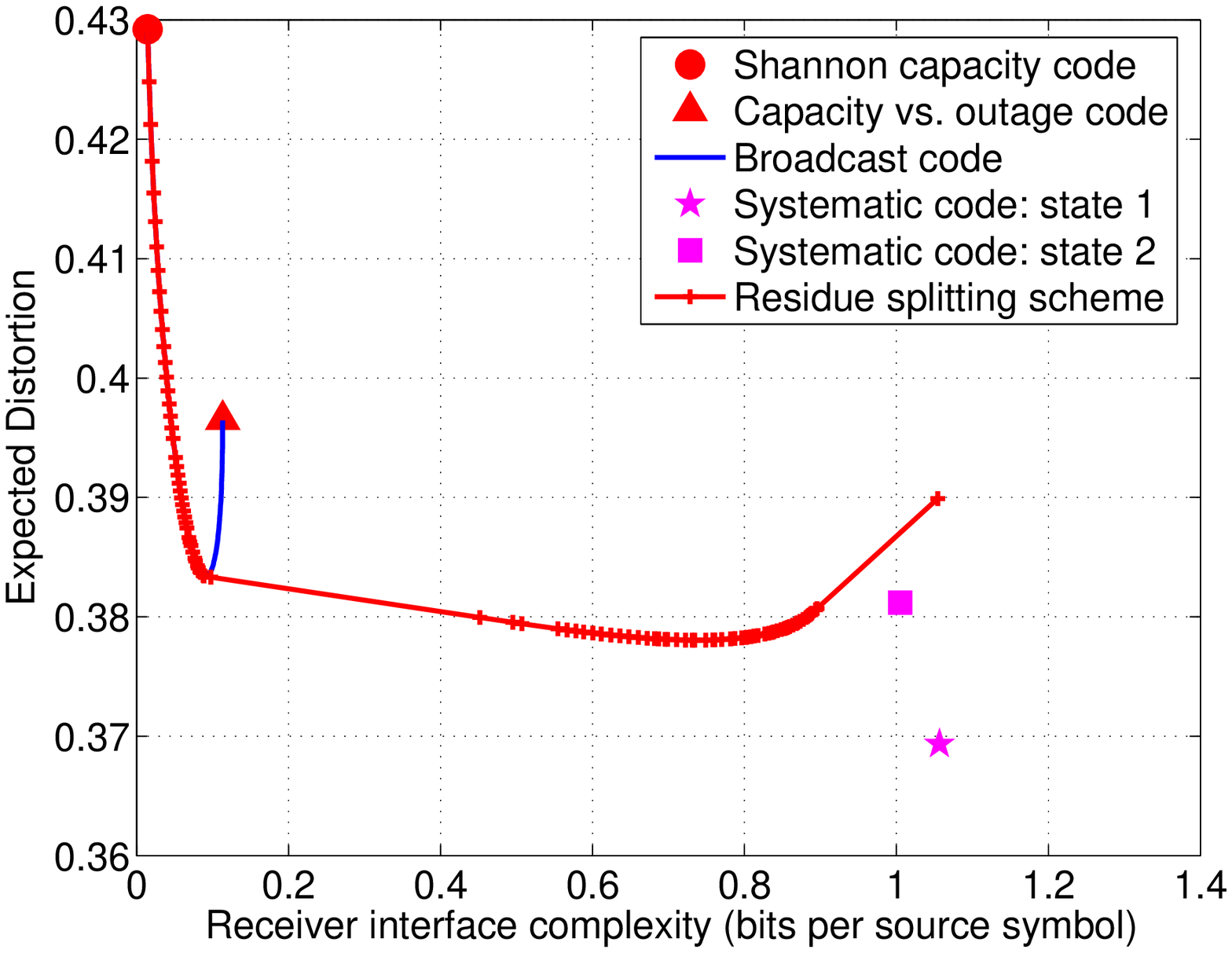}{Receiver interface
complexity vs. expected distortion tradeoff.} Expected distortion
alone does not provide the complete picture for comparison of the
schemes. In \refF{interfaceENC} and \ref{fig:interfaceDEC} we assume
the channel state probability $p = 0.7$ and illustrate the tradeoff
between the expected distortion and the transmitter/receiver
interface complexity for different schemes, where the complexity is
measured by bits per source symbol delivered through the interface.
For the broadcast scheme, we can reduce the expected distortion by
increasing $\beta$, which reduces the base layer rate but increases
the refinement layer rate and the total rate, hence a higher
interface complexity. However, the distortion-complexity curve is
not strictly decreasing. After we reach the minimum expected
distortion, it does not provide any more benefit to further increase
the interface complexity. The same trend is also observed in the
residue splitting scheme. At channel state probability $p=0.7$, the
systematic code targeting the good state has the lowest expected
distortion, nevertheless it also has the highest interface
complexity. The choice about the appropriate scheme and operating
points (parameters) depends on the system designer's view about this
distortion-complexity tradeoff.

\section{Conclusions}
\label{sec:con} We consider transmission of a stationary ergodic
source over non-ergodic composite channels with channel state
information at the receiver (CSIR). To study the source-channel
coding problem for the entire system, we include a broader class of
transmission schemes as separation schemes by relaxing the
constraint of Shannon separation, i.e. a single-number comparison
between source coding rate and channel capacity, and introducing the
concept of a source-channel interface which allows the source and
channel to agree on multiple parameters.

We show that different end-to-end distortion metrics lead to
different conclusions about separation optimality, even for the same
source and channel models. Specifically, one such generalized scheme
guarantees the separation optimality under the distortion versus
outage metric. Separation schemes are in general suboptimal under
the expected distortion metric. We study the performance enhancement
when the source and channel coders exchange more information through
a more sophisticated interface, and illustrate the tradeoff between
interface complexity and end-to-end performance through the example
of transmission of a binary symmetric source over a composite binary
symmetric channel.

\appendices
\section{MR Source Code and BC Channel Code Structure}
\label{app:SCCodeForSeparationSchemes} In \refF{expected_enc}, the
multi-resolution source code can be constructed as follows. Consider
three independent auxiliary random variables
$V_1$$\sim$Bernoulli$(\lambda)$, $V_2$$\sim$Bernoulli$(1/2)$, and
$Q_1$$\sim$Bernoulli$(D_1)$, where
\[
\lambda = \frac{D_2-D_1}{1-2D_1}
\]
and $D_1$, $D_2$ are given by \eqref{eqn:MRBCDistortionCondition}.
Also define
\[
Q_2 = V_1 \oplus Q_1,
\]
which has a Bernoulli distribution with parameter $\lambda*D_1 =
D_2$. These variables are related to the source symbol through the
relationship
\[
V = V_2 \oplus Q_2 = V_2 \oplus V_1 \oplus Q_1.
\]

{\em Random codebook generation}: Generate $2^{nbR_2}$ sequences
$V^n_2(w_2)$, $w_2 \in \{1, \cdots, 2^{nbR_2} \}$, by uniform and
independent sampling over the strong typical set $T^n_\delta(V_2)$.
Similarly, generate $2^{nbR_1}$ sequences $V^n_1(w_1)$, $w_1 \in
\{1, \cdots, 2^{nbR_1} \}$, drawn uniformly and independently over
$T^n_\delta(V_1)$.

{\em Encoding}: Given $V^n \in \mathcal{V}^n$, the encoder searches
over $(w_1,w_2) \in \{1, \cdots, 2^{nbR_1} \} \times \{1, \cdots,
2^{nbR_2} \}$. If it finds a pair $(w_1,w_2)$ such that
\[(V^n,
V^n_1(w_1), V^n_2(w_2)) \in T^n_\delta(V, V_1, V_2),
\]
it stops the search and sends the above $(w_1, w_2)$. Otherwise it
sends $(w_1,w_2)=(1,1)$.

{\em Decoding}: If only the index $w_2$ is received, the decoder
declares the estimate of the source sequence as $\hat{V}^n_2 =
V^n_2(w_2)$. If both indices are received, the source is
reconstructed as $\hat{V}^n = \hat{V}^n_1 \oplus \hat{V}^n_2 =
V^n_1(w_1) \oplus V^n_2(w_2)$. Following the procedures in
\cite{ElGamalCover82} and \cite[Theorem 1]{Tuncel03} we can easily
verify the following distortion targets are achievable:
$\mathbb{E}d(V^n, \hat{V}^n) \le D_1$, $\mathbb{E}d(V^n,
\hat{V}^n_2) \le D_2$.

In practice the MR source code can be implemented as a multi-stage
vector quantization, which has an {\em additive} successive
refinement structure \cite{Tuncel03}. As shown in
\refF{expected_enc}, in channel state 2 only the base layer
description is received and Source DEC 2 determines the base
reconstruction $\hat{V}_2^n$. When both layers are received, Source
DEC 1 determines a refinement sequence $\hat{V}_1^n$ based on the
refinement layer encoding index only, and add it to the base
reconstruction $\hat{V}_2^n$ to obtain the overall reconstruction
$\hat{V}^n$. On the contrary, for general MR source codes the
overall reconstruction may require a joint decoding of indices from
both layers. The additive refinement structure reduces coding
complexity, provides scalability, and does not incur any performance
loss under certain conditions \cite[Theorem 3]{Tuncel03}, which are
all satisfied in this example.

The broadcast channel code design, for a chosen $0 \le \beta \le
(1/2)$, is summarized as follows.

{\em Random codebook generation}: Generate $2^{nbR_2}=2^{mR_2}$
independent codewords $U^m(w_2)$, $w_2 \in \{1, \cdots, 2^{mR_2}
\}$, by i.i.d. sampling of a Bernoulli$(1/2)$ distribution. Generate
$2^{nbR_1}=2^{mR_1}$ independent codewords $Q_\beta^m(w_1)$, $w_1
\in \{1, \cdots, 2^{mR_1} \}$, by i.i.d. sampling of a
Bernoulli$(\beta)$ distribution.

{\em Encoding}: To send the index pair $(w_1,w_2)$, send $X^m =
Q_\beta^m(w_1) \oplus U^m(w_2)$.

{\em Decoding}: Given channel output $Z^m$, in state 2 we determine
the unique $\hat{\hat{w}}_2$ such that
\[
d(Z^m, U^m(\hat{\hat{w}}_2)) \le (\alpha_2 * \beta).
\]
In state 1 we look for the unique indices $(\hat{w}_1, \hat{w}_2)$
such that
\begin{eqnarray*}
&& d(Z^m, U^m(\hat{w}_2)) \le (\alpha_1 * \beta), \\
&& d(Z^m, Q_\beta^m(\hat{w}_1) \oplus U^m(\hat{w}_2)) \le \alpha_1.
\end{eqnarray*}
Following the analysis of \cite[Theorem 14.6.2]{CoverIT}, we can
show that the channel decoding error probability approaches zero as
long as the encoding rates satisfy \eqref{eqn:MRBCRateCondition}.

Roughly speaking, in channel state 2, we observe
\[
Z^m = X^m \oplus Q_{\alpha_2}^m =  U^m \oplus Q_\beta^m \oplus
Q_{\alpha_2}^m,
\]
where the channel noise $Q_{\alpha_2}^m$ is a Bernoulli$(\alpha_2)$
sequence. We want to decode the $U^m$ sequence subject to the
overall interference-plus-noise $Q_\beta^m \oplus Q_{\alpha_2}^m$,
which is a Bernoulli sequence with parameter $(\alpha_2 * \beta)$,
hence the achievable rate $1-h(\alpha_2* \beta)$. In channel state
1, we observe
\[
Z^m = X^m \oplus Q_{\alpha_1}^m =  U^m \oplus Q_\beta^m \oplus
Q_{\alpha_1}^m.
\]
Since $\alpha_1 < \alpha_2$, the sequence $U^m$ can be decoded and
then subtracted off. We then decode $Q_\beta^m$ subject to the noise
$Q_{\alpha_1}^m$, and the rate $h(\alpha_1*\beta) - h(\alpha_1)$ is
achievable.

\bibliographystyle{unsrt}

\end{document}